# Deep Sigma Point Processes for RCS Modeling in Spaceborne SAR Imagery

**Khalid El-Darymli • Christoph H. Gierull • Katerina Biron • Weimin Huang**

Defence Research and Development Canada, Ottawa, Ontario, Canada
Memorial University of Newfoundland, St. John's, Newfoundland and Labrador, Canada

### Version notice

This file is prepared as an author preprint for scholarly dissemination through arXiv.
It includes the complete article content and figures.

*Draft designation and cover page added for preprint distribution; scientific content is unchanged.*

Corresponding author: khalid.el-darymli@drdc-rddc.gc.ca

# Deep Sigma Point Processes for RCS Modeling in Spaceborne SAR Imagery

**KHALID EL-DARYMLI** , Senior Member,
**CHRISTOPH H. GIERULL** , Senior Member,
**KATERINA BIRON**
Defence Research and Development Canada, Ottawa, ON, Canada

**WEIMIN HUANG** , Senior Member,
Memorial University, St. John's, NL, Canada

**Radar cross-section (RCS) modeling is foundational to advancing the utility and sensitivity of spaceborne radar systems. This study introduces a deep sigma point process (DSPP) model for predicting RCS in synthetic aperture radar (SAR) imagery, using a RADARSAT-2 dataset containing 208 191 verified ships. The DSPP model not only strives for predictive accuracy but ventures to characterize the uncertainty inherent in the intricate relationships among radar signals, ship parameters, and environmental conditions. Unlike traditional approaches relying on deterministic equations with static parameters, the DSPP leverages a hierarchical Gaussian process framework with Bayesian inference to capture variability and uncertainty in RCS predictions. By generating predictive distributions rather than single estimates, the model effectively accounts for the complex dynamics governing radar returns. Using a Matérn kernel with automatic relevance determination, the DSPP identifies and ranks critical features across radar, operational, and environmental domains, ensuring transparency and interpretability. Performance evaluations demonstrate the model's superiority over linear regression baselines, achieving a 20.83% reduction in root mean squared error, a 25.89% increase in $R^2$, and a 44.4% reduction in both residual interquartile range and median absolute deviation on test data. By providing calibrated uncertainty bounds, the DSPP enhances prediction reliability and supports robust decision-making. This work marks a shift toward probabilistic models that incorporate the inherent uncertainty of complex phenomena. Transitioning from fixed equations to distributions over outcomes, the DSPP fosters a deeper understanding of RCS behavior, enabling systems to thrive in dynamic operational environments.**

Received 17 December 2024; revised 13 May 2025 and 9 June 2025; accepted 13 June 2025. Date of publication 24 June 2025; date of current version 13 October 2025.

Refereeing of this contribution was handled by K. S Kulpa.

Authors' addresses: Khalid El-Darymli, Christoph Gierull, and Katerina Biron are with the Department of National Defence, Defence Research and Development Canada, Ottawa, ON K1A 0Z4, Canada, E-mail: (khalid.el-darymli@drdc-rddc.gc.ca; christoph.gierull@drdc-rddc.gc.ca; katerina.biron@drdc-rddc.gc.ca); Weimin Huang is with the Department of Electrical and Computer Engineering, Memorial University of Newfoundland, St. John's, NL A1B 3X5, Canada, E-mail: (weimin@mun.ca). *(Corresponding author: Khalid El-Darymli.)*

## I. INTRODUCTION

Empirical RCS models are critical for accurately predicting the performance of spaceborne synthetic aperture radar (SAR) systems in maritime vessel detection. This is because these models enable precise estimations of radar reflectivity, which directly impacts the system's ability to detect and classify vessels in diverse environmental and operational conditions. Especially under increasingly stringent surveillance requirements, such predictions are essential for optimizing radar sensitivity and ensuring consistent detection across varying maritime scenarios. Beyond performance prediction, these models are instrumental in defining requirements for the development of new ship detection modes (SDMs). As the Department of National Defence (DND) and Canadian Armed Forces (CAF) shift focus toward detecting smaller vessels, current semiempirical models—based largely on ship length—fall short by neglecting key factors including standard and extended operating conditions [1], [2], [3], [4]. Developing more robust empirical RCS models that account for these variables is essential for improving the design of SDMs and ensuring reliable performance of future radar systems, particularly for projects, such as the defence enhanced surveillance from space (DESSP) [5].

According to the *Recommended Practice for RCS Test Procedures* Std 1502-2020), developed by the Antennas and Propagation Standards Committee, RCS, denoted by $\sigma$ and measured in square metres (m$^2$), quantifies a target's ability to reflect radar energy toward the receiver. RCS is defined as $4\pi$ times the ratio of scattered power per unit solid angle in a specified direction to the incident power per unit area. This measure represents an area that would intercept an equivalent power, radiating isotropically, to produce the same received power at the radar receiver. Mathematically, the RCS is given by [6]

$$\sigma\ [\mathrm{m}^2] = 4\pi \lim_{R\to\infty} R^2 \frac{|E_s|^2}{|E_i|^2} \tag{1}$$

where

1) $R$ is the distance from the target to the receiver;
2) $E_s$ is the scattered electric field;
3) $E_i$ is the incident electric field.

The $R \to \infty$ assumption ensures far-field (i.e., planar wave) conditions, making RCS an intrinsic target property, independent of measurement distance. This work focuses on monostatic radar (i.e., backscatter RCS), where the incident and reflected scattering directions are coincident but opposite in sense.

In practical applications, the RCS is often expressed in decibels relative to a square metre (dBsm) due to its wide dynamic range. That is,

$$\sigma \text{ [dBsm]} = 10 \log_{10} \left( \frac{\sigma \text{ [m}^2\text{]}}{1 \text{ [m}^2\text{]}} \right). \tag{2}$$

The first empirical RCS model for naval ships was developed by Merrill I. Skolnik at the Naval Research Laboratory (NRL) in the 1970 s. In his 1973 NRL report [7] and his 1974 Transactions on Aerospace and Electronic Systems paper [8], Skolnik introduced the following formula for RCS:

$$\text{RCS} = 52 f_c^{1/2} \mathcal{D}^{3/2} \tag{3}$$

where

1) RCS is in ($m^2$);
2) $f_c$ is the radar center frequency in megahertz (MHz);
3) $\mathcal{D}$ is the ship's displacement in kilotons.

This formula applies to ships in the microwave band, with RCS measurements based on naval vessels ranging from 2–17 kilotons. Skolnik's model provides a median RCS value for ship displacements at X-band (3.25-cm wavelength), S-band (10.7 cm), and L-band (23 cm). The RCS values are averaged over port and starboard bow and quarter aspects, excluding the broadside peak. As Skolnik noted in his Radar Handbook [9], [10], within classes of targets, the RCS can vary by as much as 20 to 30 dB depending on frequency, aspect angle, and target characteristics.

In a separate study [11], a marine X-band radar with a 3.2-cm wavelength and horizontally polarized transmission and reception (HH) was used to develop a verified and calibrated RCS dataset of 18 ships passing through the English Channel. These ships ranged from small fishing boats to large tankers, with gross tonnage ranging from 5–45 000 tons. Other RCS measurement ranges, such as those reported by the International Association of Marine Aids to Navigation and Lighthouse Authorities (IALA) in their 2022 report [12], confirm these measurements. Using Skolnik's formula, estimates of the median RCS for various ships showed errors ranging from 7–17 dBsm when compared to measured values, consistent with Skolnik's observation that RCS errors can reach up to 30 dBsm [3].

Further research analyzing the relationship between ship displacement and ship length using automatic identification system (AIS) data has shown that it is possible to replace the displacement parameter $\mathcal{D}$ in Skolnik's formula with ship length [13]. This relationship has been leveraged in recent works to estimate the RCS based on ship dimensions [3].

Skolnik's model has been refined over time by various researchers [14], [15], and one notable variation, introduced by Vachon et al. [1], [14], [15], [16], [17], [18], [19] and more recently by AstroCom Associates Inc [20], incorporates radar incidence angle. This empirical formula takes the form

$$\text{RCS} = a x_{\text{truthLength}}^{p} (b + c\theta) \tag{4}$$

where

1) $x_{\text{truthLength}}$ is the ship's true length in metres;
2) $\theta$ is the radar incidence angle;
3) the parameters $a$, $b$, $c$, and $p$ are inferred through fitting with the empirical data acquired at a certain center frequency, such as C-band of RADARSAT-2.

In curve-fitting literature, Skolnik's equation and its variants are referred to as power series models [21], while in machine learning, these are typically modeled using multiple linear regression [22]. In this context, "multiple" refers to the number of predictor variables, such as $f_c$ and $\mathcal{D}$ in Skolnik's model [(3)], or $x_{\text{truthLength}}$ and $\theta$ in the more recent variants [(4)]. "Linear" refers to the linear nature of the regression after appropriate transformations are applied to the data. However, RCS modeling using power series and multiple linear regression suffers from several key issues, including nonconstant variance, autocorrelation, multicollinearity, and overfitting [22]. These problems can lead to unreliable predictions and inaccurate confidence intervals.

From a broader machine learning perspective, multiple regression models are a limited special case of more advanced techniques [23], [24], [25]. These modern techniques not only address the abovementioned challenges more effectively but also they are not restricted to a $k$th percentile and provide better RCS models with tighter error bounds. Therefore, the objective of this article is to utilize recent advancements in deep learning to develop a reliable and holistic RCS model. The main innovation in this article is the introduction of a novel deep sigma point process (DSPP) model [26]. This Bayesian model not only predicts RCS values but also learns the uncertainty estimates (error bounds), enhancing its practical utility. By quantifying predictive uncertainty, the DSPP model offers a more robust approach to RCS modeling, addressing the limitations inherent in traditional regression methods. The main contributions of this work are as follows.

1) *Comprehensive Feature Utilization:* Unlike previous works that primarily focused on a limited set of features, namely, radar center frequency, target length, and radar incidence angle, this work promotes the incorporation of a wider range of relevant features. This includes not only radar parameters but also environmental variables and associated AIS parameters, among others. The inclusion of these features allows the model to characterize both standard and extended operating conditions more comprehensively.
2) *Framework Development for Model Training and Evaluation:* A robust framework was developed for training, validation, and testing of deep learning models tailored to empirical RCS modeling of targets in SAR imagery. This framework streamlines model development and benchmarking across diverse datasets and operational scenarios.
3) *Novel Bayesian DSPP Model:* This work introduces an innovative Bayesian DSPP model with the dual capability of RCS prediction and error bound estimation, enhancing prediction reliability. The model also adheres to explainable AI (XAI) principles [27], providing insight into feature relevance by generating a ranked importance of each feature. This interpretability is integral to validating model predictions in practical applications.

4) *Adaptability to Multisensor Data:* The model is designed to handle data from a wide range of radar sensors, whether operating at the same or different center frequencies, and featuring varying configurations. By incorporating center frequency, radar parameters, environmental conditions, and other critical features as inputs, the model achieves robust generalization across diverse sensor platforms, making it highly versatile for various operational scenarios.

The rest of this article is organized as follows. In Section II, we formalize the problem definition and provide an overview of regression modeling for RCS, including our novel Bayesian approach. Section III describes the AIS-verified RADARSAT-2 dataset used in this study, detailing both ship detection records and associated radar parameters, providing insights into the unique features of the dataset that enable rigorous model evaluation. Section IV outlines the baseline models, with multiple linear regression variants used for comparative analysis. Our primary contribution, the novel DSPP model, is introduced in Section V, where we elaborate on its architecture, kernel selection, and uncertainty quantification capabilities. Section VI discusses the enhanced hyperparameter optimization framework, incorporating tree-structured Parzen estimator (TPE) and asynchronous successive halving algorithm (ASHA), which significantly improves model training efficiency. In Section VII, we present a comparative evaluation of the baseline models and the DSPP, supported by visualizations and performance metrics. Finally, Section VIII concludes this article.

## II. PROBLEM DEFINITION

### A. Regression Modeling of RCS

In this article, we address the problem of predicting RCS values using regression modeling, where the relationship between a set of input features (i.e., predictor variables) $\mathbf{X}$ and the RCS values (i.e., target variable) $\mathbf{y}$ is learned. This problem can be formalized as

$$\mathbf{y} = f(\mathbf{X}) \tag{5}$$

where

1) $\mathbf{X}$ is the feature matrix of size $N \times D$;
2) $\mathbf{y}$ represents the RCS in dBsm;
3) $N$ is the number of samples;
4) $D$ represents the number of features.

Each feature vector $\mathbf{x}_i \in \mathbf{X}$ for $i = 1, \ldots, D$ is associated with either AIS parameters, radar meta-parameters, or other radar-related variables. In this study, we selected 25 features, $\mathbf{x}_1$ to $\mathbf{x}_{25}$, which are used to predict the RCS of ships from SAR images.

The RCS $\mathbf{y}$ is computed from spaceborne RADARSAT-2 SAR images, specifically from AIS-verified SAR chips. The RCS for a given ship is calculated from the SAR chip using the following formula [28]:

$$\text{RCS} = \frac{\sum \aleph^2 \Delta A}{G_{\text{ref}}^2} \tag{6}$$

where

1) $\sum \aleph^2$ is the sum of the squared Digital Numbers in the SAR chip;
2) $\Delta A$ is the pixel area in square metres;
3) $G_{\text{ref}}$ is the reference gain value used for calibration.

Equation (6) estimates RCS of an extended target by summing the calibrated backscatter power (i.e., squared pixel amplitudes) over all pixels in the detected ship region. This provides an integrated RCS estimate, which represents the aggregate radar reflectivity of the target and is widely used in operational SAR applications [19], [29]. This yields a measured RCS, which serves as the ground truth label for training and evaluating the regression model. The model aims to predict this observed RCS from a set of interpretable physical and environmental features, enabling RCS estimation under varying conditions without direct reliance on SAR imagery.

The function $f$ is the regression model that learns the relationship between the features $\mathbf{X}$ and the RCS values $\mathbf{y}$. Various forms of the regression function $f$ are used in this study, including machine learning and deep learning approaches. As baselines, we consider four variants of multiple linear regression as follows.

1) *Linear regression* [30].
2) *Ridge regression* [31].
3) *Lasso regression* [32].
4) *ElasticNet* [33].

In addition to these baseline models, we propose a novel Bayesian model based on DSPP [26], which not only predicts the RCS but also learns the error bounds for the predictions, allowing for uncertainty quantification.

### B. Splitting and Normalizing the Dataset

To ensure that the trained model generalizes well, the dataset is divided into three splits: 1) training (70%); 2) validation (15%); and 3) testing (15%). The training split is used to learn $f$, while the validation split is used to tune hyperparameters and prevent overfitting. Once the model is trained, it is evaluated on the test split to assess its real-world performance. The dataset is randomly shuffled before being split. Random shuffling is performed to ensure that the data are distributed evenly across the splits, reducing the risk of bias and ensuring that the training, validation, and testing sets contain diverse samples [34].

In this study, both the input features $\mathbf{X}$ and the target variable $\mathbf{y}$ are normalized to ensure that they are on the same scale. More precisely, each feature vector $\mathbf{x}_i$ is normalized to the same range $\in [-1, 1]$ as follows:

$$\mathbf{x}_i^{\text{norm}} = -1 + 2\frac{\mathbf{x}_i - \min(\mathbf{x}_i)}{\max(\mathbf{x}_i) - \min(\mathbf{x}_i)} \tag{7}$$

where

1) $\min(\mathbf{x}_i)$ is the minimum value for the feature samples of the training split;
2) $\max(\mathbf{x}_i)$ is the maximum value for the feature samples of the training split.

This scaling helps improve convergence and performance in machine learning models by ensuring that features are comparable in magnitude, preventing any one feature from disproportionately influencing the model. Similarly, the target variable $\mathbf{y}$ (i.e., RCS in dBsm) is normalized [35]

$$\mathbf{y}^{\text{norm}} = \frac{\mathbf{y} - \mu_{\mathbf{y}}}{\sigma_{\mathbf{y}}} \tag{8}$$

where

1) $\mu_{\mathbf{y}}$ is the mean for the target variable across all samples in the training split of the dataset;
2) $\sigma_{\mathbf{y}}$ is the standard deviation for the target variable across all samples in the training split of the dataset.

## C. Performance Metrics

The performance of models is evaluated using two key metrics as follows.

1) *Root Mean Squared Error (RMSE)*: This measures the average magnitude of the error between the predicted (i.e., $\hat{y}_i$) and true (i.e., $y_i$) RCS values [36]

$$\text{RMSE} = \sqrt{\frac{1}{N}\sum_{i=1}^{N}(y_i - \hat{y}_i)^2}. \tag{9}$$

1) *Range of RMSE:* RMSE $\geq 0$;
2) A value of 0 indicates perfect predictions. Higher values indicate larger deviations between predicted and true values;
3) The unit of RMSE is the same as $y_i$;
4) *Interpretation:* Lower RMSE values are better, indicating more accurate predictions.

2) *$R^2$ Score (Coefficient of Determination):* This measures the proportion of variance in $\mathbf{y}$ that is explained by $\mathbf{X}$ [37]

$$R^2 = 1 - \frac{\sum_{i=1}^{N}(y_i - \hat{y}_i)^2}{\sum_{i=1}^{N}(y_i - \bar{y})^2} \tag{10}$$

where $\bar{y}$ is the mean of the true values $y_i$

1) *Range of $R^2$:* $-\infty < R^2 \leq 1$;
2) An $R^2 = 1$ indicates perfect prediction, while $R^2 = 0$ means no variance is explained;
3) Negative values indicate that the model performs worse than predicting the mean;
4) *Interpretation:* Higher $R^2$ values are better, indicating that more variance in RCS is captured by the model.

To calculate RMSE and $R^2$ for unnormalized data, the following conversions are used.

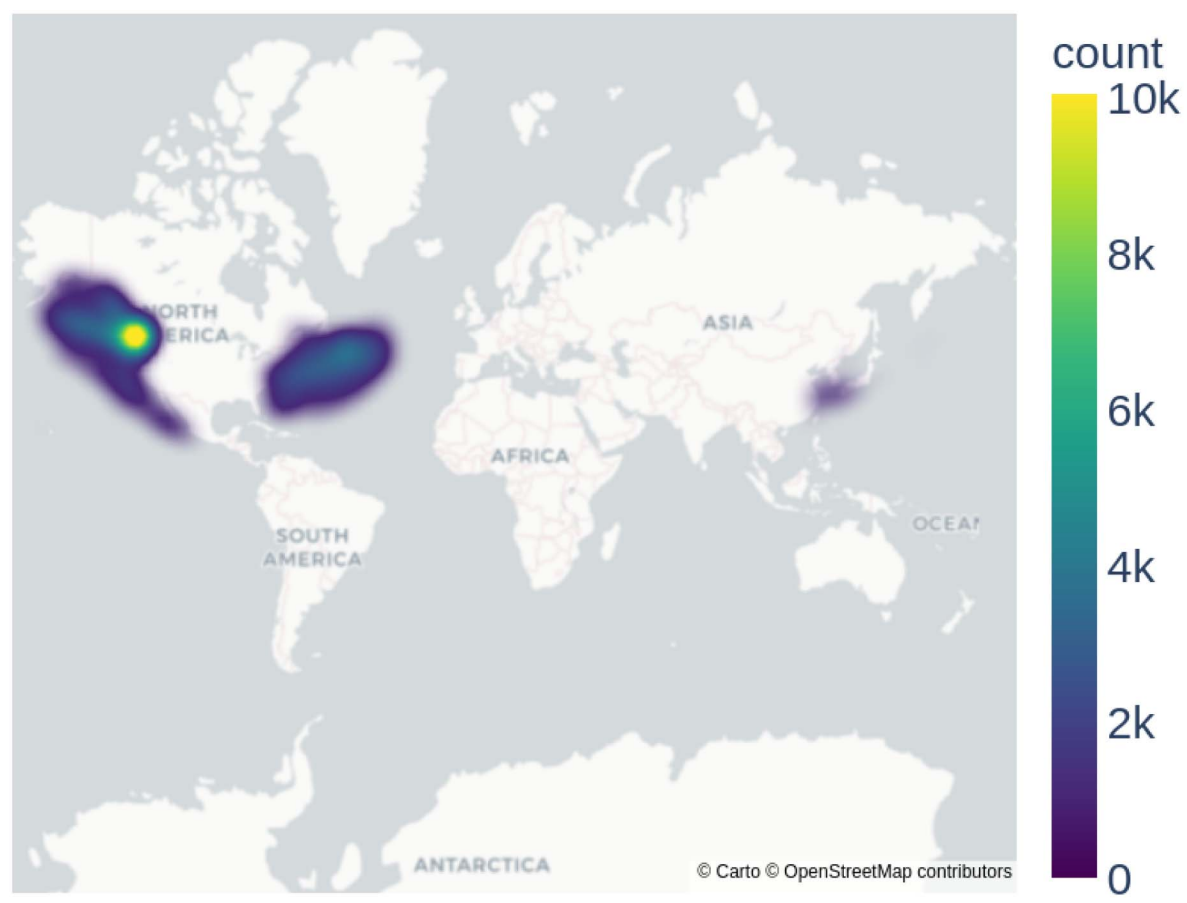

Fig. 1. Geographical distribution of the 208,191 AIS-verified RADARSAT-2 ships used in this study.

1) *Converting RMSE to Unnormalized Scale:* Once the RMSE is calculated for the normalized data ($\text{RMSE}_{\text{norm}}$), the corresponding RMSE for unnormalized data ($\text{RMSE}_{\text{unnorm}}$) can be obtained by multiplying it by the standard deviation ($\sigma_y$) of the original target variable $y$

$$\text{RMSE}_{\text{unnorm}} = \text{RMSE}_{\text{norm}} \cdot \sigma_y. \tag{11}$$

2) *Converting $R^2$ to Unnormalized Scale:* The $R^2$ value remains the same for both normalized and unnormalized data because it measures the proportion of variance explained by the model. Therefore

$$R^2_{\text{unnorm}} = R^2_{\text{norm}}. \tag{12}$$

These performance metrics provide insight into how well the model predicts the RCS values, and the unnormalized RMSE gives an accurate understanding of the prediction error in real-world units (dBsm for RCS).

# III. AIS-VERIFIED RADARSAT-2 DATASET

This study is built on and evaluated using focused, detected SAR intensity products that have undergone stationary-world matched filtering—the same magnitude-only imagery routinely used in operational maritime surveillance. These data feed processors, such as OceanSuite, the DND framework for ship detection [19], and SUMO, the open-source SAR-based ship detection module maintained by the European Commission's Joint Research Centre (JRC) [29].

The spaceborne SAR dataset used in this study was acquired between January 2013 to May 2019 and it comprises 208 191 AIS-verified ship detections, offering a comprehensive overview of maritime vessel activity captured via RADARSAT-2. The geographical distribution of these detections is represented in Fig. 1, which highlights areas of varying vessel densities across the globe. High-density clusters are primarily found along the east coast of North America, extending from the northeastern United States to southeastern Canada, and along major Pacific Ocean shipping routes near the west coast of Central America.

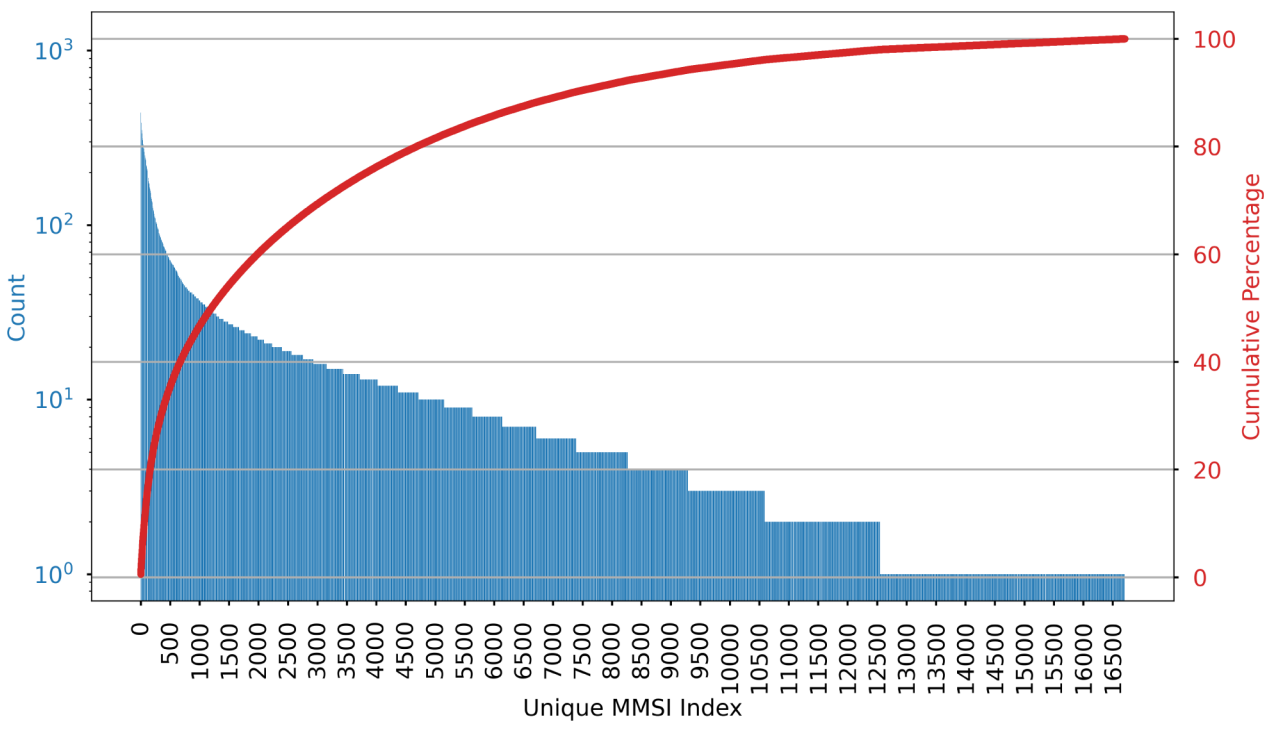

Fig. 2. Count of unique MMSIs for the AIS-verified RADARSAT-2 ships used in this study. Each tick on the x-axis represents a unique MMSI (anonymized index), and the *y*-axis shows the number of AIS-verified RADARSAT-2 detections recorded for that vessel.

A notable cluster is also observed in Southeast Asia, reflecting important regional maritime routes.

Each SAR chip in the dataset is complemented by meta files that provide a wealth of information offering a comprehensive perspective on each vessel's detection scenario, including the following.

1) The RCS in dBsm, calculated from the SAR chip using the sum of the squared digital numbers and scaled by the area of the radar chip.
2) Radar parameters, such as polarization, beam mode, and incidence angle.
3) AIS data like vessel identification and dimensions.
4) Environmental data, such as wind speed and direction, at the time of detection.

A more detailed description of the dataset can be found in [38]. A histogram of unique maritime mobile service identity (MMSI) counts is shown in Fig. 2, highlighting the distribution of vessel encounters in the dataset. The distribution demonstrates a long-tail pattern, where a small number of vessels are frequently detected, while the majority are detected less frequently. This is crucial for understanding the variability in the RCS values across different ships.

In this study, the feature matrix $\mathbf{X}$ consists of 25 features that provide detailed information about the SAR detection, radar parameters, and AIS data. The target variable $\mathbf{y}$ is the RCS, which represents the radar signal's reflected power from a ship in dBsm. The feature matrix $\mathbf{X}$ has dimensions $N \times D$, where $N = 208191$ and $D = 25$. It should be noted that these features are not exclusive and in future works additional features can be incorporated, for example by including the center frequency when considering datasets with different carrier frequencies. 2-D histogram plots for all the dataset samples considered in this study, pertaining to RCS values plotted against their corresponding features $\mathbf{x_1}$ to $\mathbf{x_{25}}$ are provided in Figs. 3 and 4. These features are defined as follows.

1) *ChipPolarization* ($\mathbf{x}_1$): The transmit/receive polarization of the SAR chip ∈ {HH, HV, VH, VV}. H and V, respectively, represent horizontal and vertical polarization. There are 191 628 chips in the dataset that are HH-polarized, while 16 563 chips are VV-polarized. There are neither HV nor VH-polarized chips in the dataset. In the feature vector $\mathbf{x}_1$, the polarization labels HH and VV, respectively, are encoded to 1 and 2.
2) *SatelliteTrackDegrees* ($\mathbf{x}_2$): The angle in degrees (from North) at which the satellite tracks along the Earth's surface.
3) *SatelliteLookDirectionDegrees* ($\mathbf{x}_3$): The direction in degrees (from North) the sensor is pointing (i.e., for default right looking configuration this is equal to SatelliteTrack+90 degrees).
4) *BeamMode* ($\mathbf{x}_4$): The beam mode in the dataset ∈ { W,[1] FW,[2] EH,[3] F,[4] UW,[5] SCN,[6] SCW,[7] DVWF,[8] and OSVN[9] } [39]. In the feature vector $\mathbf{x_4}$, the beam mode labels are encoded from 1 to 9, respectively, and the corresponding number of chips per each beam mode ∈ {237, 1769, 10, 34, 1, 5658, 797, 63717, 135968}
5) *NumberOfBeams* ($\mathbf{x}_5$): The product number of beams ∈ {1, 2, 3, 4, 7, 8}, respectively, with a corresponding number of chip samples ∈ {2079, 5457, 201, 797, 63717, 135940}.
6) *ProductNumberOfPolarizations* ($\mathbf{x}_6$): The product number of polarizations ∈ {1, 2}, with a corresponding number of chip samples ∈ {144380, 63811}, respectively.
7) *ProductNumberOfRangeLooks* ($\mathbf{x}_7$): The product number of independent range looks ∈ {1, 2, 4, 5}, with a corresponding number of chip samples ∈ {2079, 5658, 63717, 136737}, respectively.
8) *ProductNumberOfAzimuthLooks* ($\mathbf{x}_8$): The product number of independent azimuth looks ∈ {1, 2, 4}, with a corresponding number of chip samples ∈ {201489, 6455, 247}, respectively.
9) *ImageProjection* ($\mathbf{x}_9$): Projection of the SAR chip ∈ {ground-range, slant-range}. In this study, all the chips are in the ground-range projection.
10) *LineSpacingMetres* ($\mathbf{x}_{10}$): The product line spacing in metres ∈ {1.5625, 5, 6.25, 12.5, 20, 25, 50}, with a corresponding number of chip samples ∈ {1, 28, 1803, 247, 63717, 141598, 797}, respectively.
11) *PixelSpacingMetres* ($\mathbf{x}_{11}$): The product pixel spacing in metres ∈ {1.5625, 5, 6.25, 12.5, 20, 25, 35, 50}, with a corresponding number of chip samples

[1] Wide
[2] Wide Fine
[3] Extended High
[4] Fine
[5] Wide Ultra Fine
[6] ScanSAR Narrow
[7] ScanSAR Wide
[8] Detection of Vessels, Wide swath, Far incidence angle
[9] Ocean Surveillance, Very wide swath, Near incidence

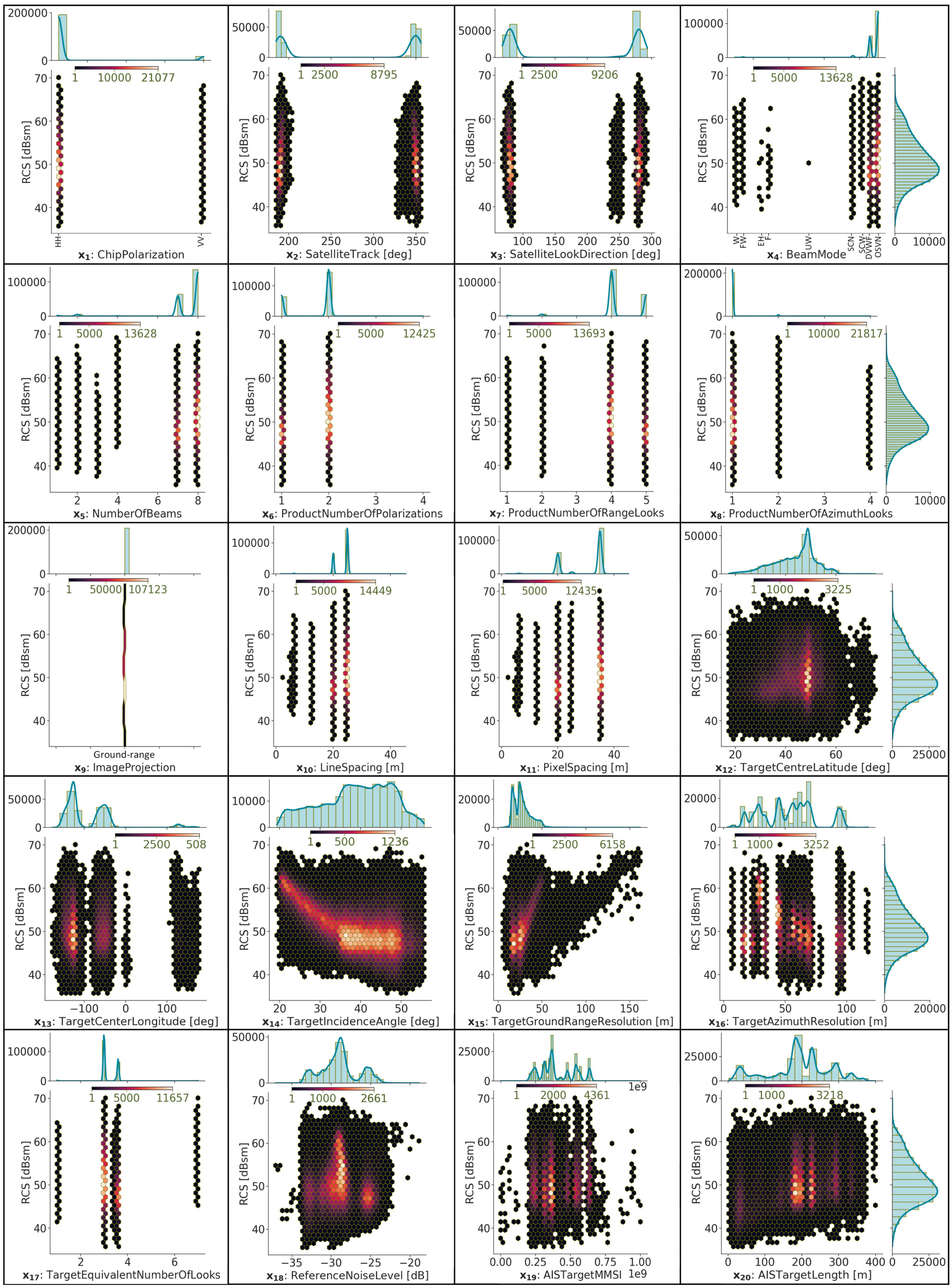

Fig. 3. 2-D histogram plots of RCS versus features $\mathbf{x_1}$ to $\mathbf{x_{20}}$.

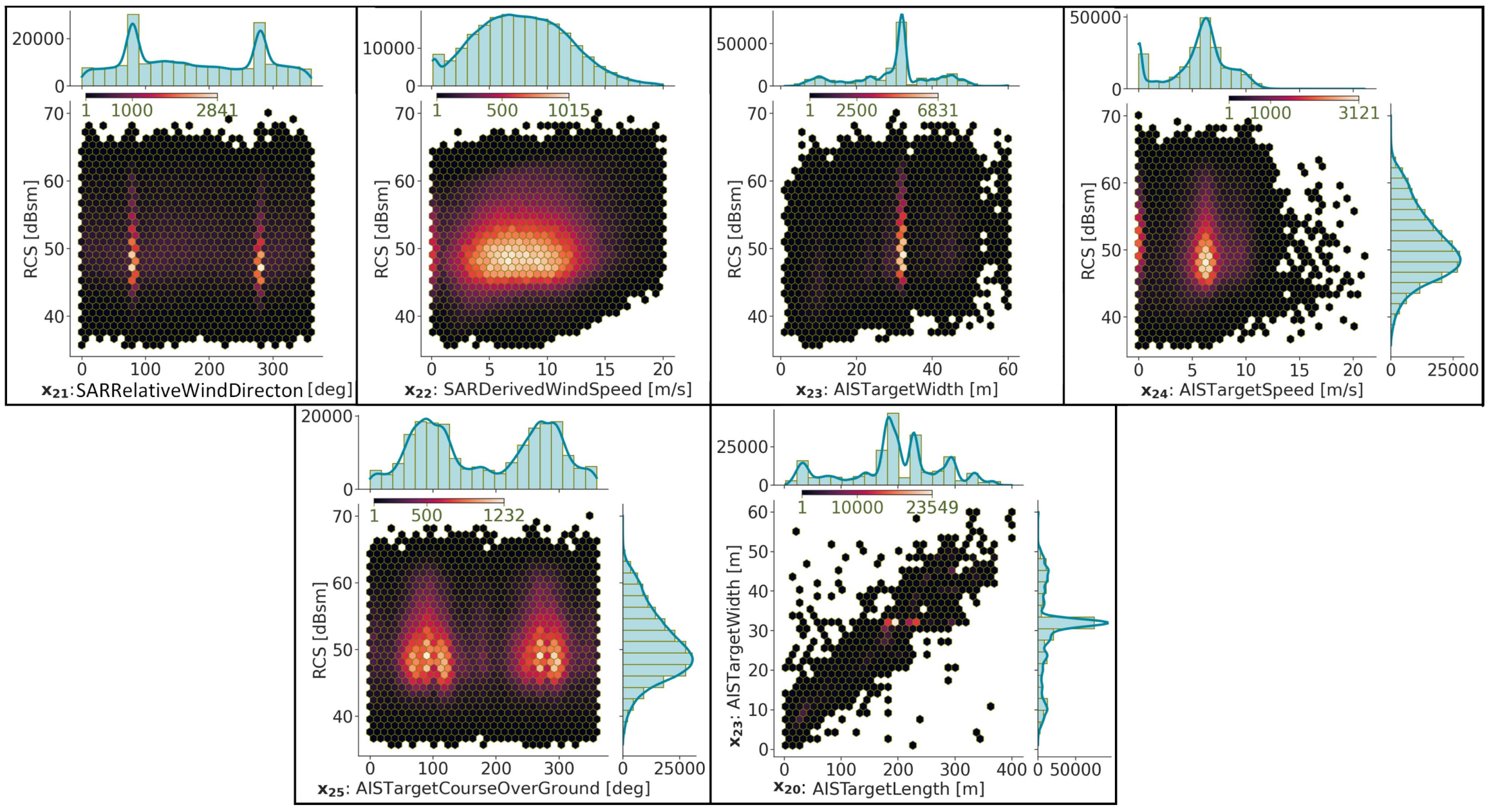

Fig. 4. 2-D histogram plots of RCS versus features $\mathbf{x}_{21}$ to $\mathbf{x}_{25}$, and $\mathbf{x}_{23}$ versus $\mathbf{x}_{20}$.

$\in \{1, 28, 1803, 247, 63717, 5658, 135940, 797\}$, respectively.

12) *TargetCenterLatitudeDegrees* ($\mathbf{x}_{12}$): The latitude of the target's center in the SAR image in degrees.
13) *TargetCenterLongitudeDegrees* ($\mathbf{x}_{13}$): The longitude of the target's center in the SAR image in degrees.
14) *TargetIncidenceAngleDegrees* ($\mathbf{x}_{14}$): The radar incidence angle at target in degrees.
15) *TargetGroundRangeResolutionMetres* ($\mathbf{x}_{15}$): The ground-range resolution of the SAR chip in metres.
16) *TargetAzimuthResolutionMetres* ($\mathbf{x}_{16}$): The azimuth resolution of the SAR chip in metres.
17) *TargetEquivalentNumberOfLooks* ($\mathbf{x}_{17}$): The equivalent number of looks for the SAR chip.
18) *ReferenceNoiseLeveldB* ($\mathbf{x}_{18}$): The reference noise level in dB at the target in the SAR chip.
19) *AISTargetMMSI* ($\mathbf{x}_{19}$): The AIS MMSI identifier of the verified ship in the SAR image.
20) *AISTargetLengthMetres* ($\mathbf{x}_{20}$): The AIS length in metres for the verified ship in the SAR image.
21) *SARRelativeWindDirectionDegrees* ($\mathbf{x}_{21}$): The relative Canadian Meteorological Centre (CMC) wind direction in degrees for the verified ship in the SAR image.
22) *SARDerivedWindSpeedMPS* ($\mathbf{x}_{22}$): The wind speed in m/s derived from the SAR product.
23) *AISTargetWidthMetres* ($\mathbf{x}_{23}$): The AIS width in metres for the verified ship in the SAR image.
24) *AISTargetSpeedMPS* ($\mathbf{x}_{24}$): The AIS speed in m/s for the verified ship in the SAR image.
25) *AISTargetCourseOverGroundDegrees* ($\mathbf{x}_{25}$): The AIS course over ground in degrees for the verified ship in the SAR image.

## IV. Baseline Models: Multiple Linear Regression Variants

To provide a robust comparison against our proposed model, we have employed four baseline multiple linear regression models, namely, *linear regression*, *ridge regression*, *lasso regression*, and *ElasticNet*. Each model applies different regularization techniques to improve the generalization of the linear regression model, especially when dealing with high-dimensional or correlated data.

### A. Linear Regression

Linear regression is a fundamental model that estimates the relationship between a dependent variable $\mathbf{y}$ and independent variables $\mathbf{X}$ by minimizing the residual sum of squares between observed values and the values predicted by the linear approximation [30]. The mathematical form of linear regression is as follows:

$$\hat{\mathbf{y}} = \mathbf{X}\boldsymbol{\beta} + \boldsymbol{\epsilon} \tag{13}$$

where $\mathbf{X} \in \mathbb{R}^{N\times D}$ represents the input features, $\beta \in \mathbb{R}^{D}$ are the model parameters, and $\epsilon \in \mathbb{R}^{N}$ denotes the error term. The goal of the model is to find $\hat{\boldsymbol{\beta}}$ that minimizes the following cost function:

$$\min_{\boldsymbol{\beta}} \|\mathbf{y} - \mathbf{X}\boldsymbol{\beta}\|_2^2. \tag{14}$$

Linear regression assumes that the relationship between the input features and the output variable is linear, and no regularization is applied to the model coefficients.

### B. Ridge Regression

Ridge regression extends linear regression by introducing an $L_2$ regularization term to the cost function, which penalizes large model coefficients. This helps in preventing overfitting, particularly when multicollinearity exists in the data [31]. The ridge regression optimization problem is formulated as

$$\min_{\boldsymbol{\beta}} \|\mathbf{y} - \mathbf{X}\boldsymbol{\beta}\|_2^2 + \lambda\|\boldsymbol{\beta}\|_2^2 \tag{15}$$

where $\lambda \geq 0$ is the regularization parameter that controls the strength of the penalty. As $\lambda$ increases, the magnitude of the model coefficients is shrunk, which can reduce variance but may introduce bias.

### C. Lasso Regression

Lasso (least absolute shrinkage and selection operator) regression applies $L_1$ regularization to the model coefficients, encouraging sparsity by driving some coefficients to zero. This makes lasso particularly useful for feature selection [32]. The optimization problem for lasso regression is given by

$$\min_{\boldsymbol{\beta}} \|\mathbf{y} - \mathbf{X}\boldsymbol{\beta}\|_2^2 + \lambda\|\boldsymbol{\beta}\|_1. \tag{16}$$

Here, $\|\boldsymbol{\beta}\|_1$ represents the $L_1$ norm, and $\lambda \geq 0$ controls the degree of regularization. Unlike ridge, lasso can produce sparse solutions by forcing some of the coefficients to be exactly zero, effectively selecting a subset of the features.

### D. ElasticNet

ElasticNet combines both $L_1$ and $L_2$ regularization techniques by linearly combining the penalties of ridge and lasso. This allows the model to benefit from the sparsity of lasso and the stability of ridge, making it suitable for cases where the number of predictors exceeds the number of observations, or when there are highly correlated features [33]. The optimization problem for ElasticNet is as follows:

$$\min_{\boldsymbol{\beta}} \|\mathbf{y} - \mathbf{X}\boldsymbol{\beta}\|_2^2 + \lambda_1\|\boldsymbol{\beta}\|_1 + \lambda_2\|\boldsymbol{\beta}\|_2^2 \tag{17}$$

where $\lambda_1$ and $\lambda_2$ are the regularization parameters for the $L_1$ and $L_2$ penalties, respectively. ElasticNet is useful when neither ridge nor lasso alone provides an adequate model for the data.

## V. NOVEL MODEL – DEEP SIGMA POINT PROCESS

DSPP is a hierarchical Gaussian process (GP) model designed to capture complex, nonlinear relationships in data while estimating predictive uncertainty. As illustrated in Fig. 12, the DSPP architecture comprises multiple layers where input features flow through several GPs, each transforming the data while propagating uncertainty. This section discusses the model's architecture, including the roles of sigma points, kernels, inducing points, and quadrature rules, and how dimensionality changes throughout the process.

### A. Summary of Key Variables

1) $W$: The width of each GP layer, representing the number of GPs in that layer.
2) $L$: The number of layers in the DSPP model, where each layer processes the outputs from the previous layer.
3) $M$: The number of inducing points used to approximate the GP posterior for large datasets.
4) $S$: The number of sigma points (quadrature points) used to approximate the integral over latent functions.
5) $\xi_w^{(s)}$: Sigma points that approximate the latent variable distribution, transforming the integral into a finite mixture.
6) *Kernel Function* $k(x_i, x_j)$*:* Defines the covariance between input points and determines how outputs are related in each GP. The kernel is applied separately to each GP node at each layer, with various kernels selectable based on data characteristics.

### B. Why DSPP?

DSPP offers a robust framework for modeling nonlinear relationships while estimating mean predictions and their associated uncertainties. This capability is crucial in fields that require precise uncertainty assessments. As a Bayesian model, DSPP provides distributions over predictions rather than just point estimates, facilitating the quantification of uncertainty [26]. Key advantages of DSPP include the following.

1) *Mean and Variance Estimation:* DSPP predicts both the mean and variance of outputs, providing valuable error bounds for risk assessment.
2) *Uncertainty Propagation:* The model captures how uncertainty evolves through multiple GP layers, allowing for the assessment of prediction reliability.

DSPP excels at propagating uncertainty across layers of GPs, enabling it to:

1) *Identify Risk:* confidence intervals provided by DSPP clarify prediction reliability, enhancing decision-making processes;
2) *Model Uncertainty in Complex Systems:* DSPP effectively handles noisy, sparse data from intricate systems, yielding predictions with error bounds that reflect inherent variability.

### C. Why Use Sigma Points? the Link to Kalman Filtering and Other Fields

Sigma points in DSPP draw inspiration from their application in unscented Kalman filtering (UKF) and other domains where uncertainty propagation is essential. In Kalman filtering, particularly in control systems and robotics, the challenge lies in propagating uncertainty through nonlinear systems. UKF utilizes sigma points to approximate the distribution of state variables, capturing the mean and covariance without relying on stochastic sampling, thus enabling efficient updates of estimates [40]. In DSPP, sigma points serve similar functions as follows.

1) *Deterministic Quadrature Points:* They approximate integrals over latent functions in GPs, effectively capturing both mean and variability.
2) *Efficient Uncertainty Propagation:* DSPP leverages sigma points to propagate uncertainty across multiple GP layers, avoiding the computational costs associated with Monte Carlo sampling.

Sigma points also find applications beyond Kalman filtering as follows.

1) *Control Systems:* They approximate uncertainty in evolving system states [40].
2) *Sensor Fusion:* Sigma points facilitate data fusion from multiple sensors by propagating uncertainty [41].
3) *Robotics:* They are used to track objects and navigate autonomous systems, where noise and uncertainty must be carefully managed [42], [43].

### D. Input Layer and Dimensionality

The input to the DSPP model consists of feature vectors $\mathbf{x} = [x_1, x_2, \ldots, x_D]$, where $D$ denotes the number of features per sample. For a dataset with $N$ samples, the input data is represented as a matrix $\mathbf{X} \in \mathbb{R}^{N \times D}$. As the input features progress through multiple layers of GPs, they are transformed into latent representations, with dimensionality evolving throughout the hierarchy. In the first GP layer, the input features are processed by $W$ independent GPs, where $W$ represents the width of the layer. Each GP computes a latent variable with a multivariate Gaussian distribution $\mathcal{N}$ for each input sample $\mathbf{x}_i$, characterized by a mean $\mu_{g_w}(\mathbf{x}_i)$ and variance $\sigma^2_{g_w}(\mathbf{x}_i)$

$$g_w(\mathbf{x}_i) \sim \mathcal{N}(\mu_{g_w}(\mathbf{x}_i), \sigma^2_{g_w}(\mathbf{x}_i)), \quad w = 1, 2, \ldots, W. \quad (18)$$

### E. Role of the Kernel

The kernel function $k(x_i, x_j)$ defines the covariance structure between input points, playing a crucial role in determining correlations among different GPs based on input data. Specifically, the kernel function:

1) *Defines Covariance:* it computes the covariance matrix $\mathbf{K}$, with each entry $K_{ij} = k(x_i, x_j)$ representing the covariance between two inputs;
2) *Governs Uncertainty Propagation:* it shapes how uncertainty is propagated across layers by defining correlations between latent variables.

Each GP utilizes its own kernel function, with various types available (e.g., radial basis function, Matérn, periodic, etc.) depending on the data characteristics [3]. In addition, additive and multiplicative composites of these kernels can be employed to capture more complex patterns in the data.

### F. Inducing Points in DSPP

Computing and inverting large covariance matrices in GPs can be computationally expensive for large datasets (complexity $\mathcal{O}(N^3)$). To enhance scalability, DSPP introduces inducing points $M$. These points approximate the GP's latent functions, allowing the model to reduce computational complexity to $\mathcal{O}(NM^2)$ while still capturing essential GP structures. Inducing points approximate the covariance between latent outputs and input data points at each GP layer. The kernel function computes the covariance between the inducing points and input features, facilitating efficient uncertainty propagation

$$\mathrm{Cov}[f(\mathbf{z}), f(\mathbf{x})] = k(\mathbf{z}, \mathbf{x}) \quad (19)$$

where $\mathbf{z} \in \mathbb{R}^{M \times D}$ are the inducing points, and $\mathbf{x}$ are the input data points.

### G. Sigma Points

Sigma points (denoted as $\xi_w^{(s)}$) approximate intractable integrals over latent variables in DSPP. Serving as deterministic quadrature points, they enable efficient approximation of nonlinear transformations of latent functions without the need for stochastic sampling, thereby propagating both the mean and uncertainty through the GP layers. For each GP, the latent output $g_w(\mathbf{x}_i)$ is approximated using a weighted combination of the mean and variance

$$g_w(\mathbf{x}_i) = \mu_{g_w}(\mathbf{x}_i) + \xi_w^{(s)} \sigma_{g_w}(\mathbf{x}_i), \quad s = 1, 2, \ldots, S \quad (20)$$

where

1) $S$ is the total number of sigma points;
2) $\xi_w^{(s)}$ are the learnable sigma points capturing variability and uncertainty.

Sigma points effectively approximate the integrals needed to propagate uncertainty across GP layers.

### H. Quadrature Rules in DSPP

Quadrature rules are techniques for approximating integrals when exact computation is intractable. In DSPP, quadrature rules are implemented via sigma points, allowing efficient approximation of integrals over latent variable distributions. Instead of calculating exact integrals over latent functions, DSPP employs sigma points as discrete representations of latent variable distributions, forming a weighted sum that approximates the integral [26]

$$\int p(f)\,df \approx \sum_{s=1}^{S} w_s p(\xi_s) \quad (21)$$

where $w_s$ are the weights associated with each sigma point. This transforms the integral into a manageable sum, facilitating efficient computation while capturing necessary uncertainty.

Utilizing quadrature rules through sigma points enables DSPP to maintain high prediction accuracy while avoiding the computational burden of traditional sampling methods, making it suitable for high-dimensional inputs and large datasets.

### I. Hidden GP Layers

DSPP consists of $L$ layers of GPs, each with $W$ GPs (nodes). In each layer, inducing points approximate the latent function, while sigma points propagate uncertainty. For instance, in Fig. 12, the DSPP model features parameters $L = 2$, $W = 4$, and $S = 10$, serving as a reference for arbitrary configurations. For each GP at layer $l$, the latent output is computed based on the kernel function

$$f_w^{(l)}(g_w^{(l-1)}(\mathbf{x})) \sim \mathcal{N}\left(\mu_{f_w^{(l)}}(g_w^{(l-1)}(\mathbf{x})), \sigma^2_{f_w^{(l)}}(g_w^{(l-1)}(\mathbf{x}))\right), \quad w = 1, 2, \ldots, W. \tag{22}$$

The kernel function determines the covariance between inputs and inducing points at each layer, while inducing points help reduce computational complexity and sigma points approximate integrals over latent variables.

### J. Predictive Distribution

The predictive distribution for the target $y_i$ given the input $\mathbf{x}_i$ involves both multiplication (due to the hierarchical GP structure) and integration (to account for uncertainty over the latent variables)

$$p(y_i|\mathbf{x}_i) = \int \prod_{l=1}^{L} p(y_i|f^{(L)}(\mathbf{x}_i)) \prod_{l=1}^{L} p(f^{(l)}|f^{(l-1)}, \mathbf{x}_i) \, df^{(1)} \ldots df^{(L)}. \tag{23}$$

This integral is approximated using sigma points

$$p_{\text{DSPP}}(y_i|\mathbf{x}_i) \approx \sum_{s_1=1}^{S} \cdots \sum_{s_W=1}^{S} \omega_{s_1,..,s_W} \mathcal{N}(y_i|\mu_f^{(L)}(g_w^{(L-1)}(\mathbf{x}_i)), \sigma_f^{(L)}(g_w^{(L-1)}(\mathbf{x}_i)) + \sigma^2_{\text{obs}}) \tag{24}$$

where

1) $\mu_f^{(L)}$ and $\sigma_f^{(L)}$ are the mean and variance from the final GP layer;
2) $\omega_{s_1,\ldots,s_W}$ are the mixture weights for each combination of sigma points;
3) $\sigma_{\text{obs}}$ is the variance of the Normal likekihood $p(y|.)$.

The kernel function computes the covariance between inputs and inducing points, while sigma points propagate uncertainty through the GPs.

### K. Loss Function

The loss function for DSPP consists of a log-likelihood term that evaluates the fit of the model to the data and a regularization term that penalizes Kullback–Leibler (KL) divergence from the GP prior [26]

$$\mathcal{L}_{\text{DSPP}} = \sum_{i=1}^{N} \log p_{\text{DSPP}}(y_i|\mathbf{x}_i) - \beta_{\text{reg}} \sum \text{KL} \tag{25}$$

where:

1) *Log-Likelihood:* The log of the predictive distribution (approximated using sigma points and kernels) is computed for each data point.
2) *Regularization (KL Divergence):* A regularization term $\beta_{\text{reg}}$ that penalizes the KL divergence between the posterior distribution over inducing points and their GP prior, ensuring smoothness and preventing overfitting.

### L. Training the DSPP Model

Training the DSPP model typically involves using optimization algorithms, such as stochastic gradient descent (SGD) or Adam to minimize the loss function over a specified number of epochs. The choice of epochs and learning rate is critical for convergence and model performance.

1) *Epochs:* The number of epochs required for training depends on the complexity of the data and model architecture. Monitoring the validation loss is essential to avoid overfitting and ensure that the model generalizes well to unseen data.
2) *Optimization Algorithm:* Adam, known for its adaptive learning rates, is often preferred due to its effectiveness in handling sparse gradients and faster convergence, while SGD can be beneficial for larger datasets requiring lower memory overhead. RMSProp is closely related to Adam [44].

By incorporating inducing points, sigma points, quadrature rules, and kernel functions, DSPP efficiently approximates the propagation of both mean and uncertainty across layers, enabling scalable and accurate modeling of complex, nonlinear data. The kernels define the covariance structure at each GP, inducing points reduce computational costs, and sigma points transform intractable integrals into manageable sums, allowing DSPP to handle large datasets and high-dimensional inputs effectively. This Bayesian framework enhances predictive capabilities and empowers practitioners with tools to assess the reliability and risk associated with their predictions.

## VI. ENHANCED HYPERPARAMETER OPTIMIZATATION USING TPE AND ASHA

Hyperparameter optimization plays a critical role in machine learning, as it significantly impacts model performance. Traditional methods, such as grid and random searches are often inefficient, requiring substantial computational resources and time. Recent advances in probabilistic and heuristic approaches have demonstrated significant potential for improving scalability and performance in hyperparameter optimization tasks [45]. The TPE, introduced by Bergstra et al. [46], is a Bayesian optimization algorithm that models the likelihood of hyperparameter configurations yielding lower losses. Unlike traditional methods, which search the hyperparameter space uniformly, TPE dynamically adjusts its search based on historical performance, focusing on regions of the search space that are more likely to result in better model performance. By

**Algorithm 1:** Enhanced Hyperparameter Optimization using TPE and ASHA.

```
1   Input: Hyperparameter search space
    Input: Budget constraints: n_max_trials = 512,
           CPUs = 16, GPUs = 0.5
    Initialize: TPE with n_startup_trials = 64,
                seed = 42
    Initialize: ASHA with n_epochs = 4096,
                grace_period = 16,
                reduction_factor = 2
2   while trials ≤ n_max_trials and resources available do
3       Hyperparameter Suggestion by TPE:
          Suggest new hyperparameters based on
          previous trial performance;
4       Trial Execution: Start a new trial with the
          suggested hyperparameters, using
          allocated CPUs and GPUs;
5       Monitor "val_rmse" or "val_loss' metric to
          evaluate performance;
6       if trial performance improves significantly
          then
7           Increase resource allocation
              proportionally;
8       else
9           Capture and log any errors or unusual
              behavior;
10          if minimal improvement observed over
              defined iterations then
11              Terminate trial early;
12      if performance of "val_rmse" or "val_loss"
          is promising then
13          Continue and allocate additional
              resources;
14      else
15          Terminate trial early to conserve
              resources;
16      Update TPE Model: Feed trial results back
          to TPE to refine hyperparameter model;
17  Output: Identify and output best-performing
            hyperparameters based on lowest
            "val_rmse" or "val_loss"
18  Output: Save state of optimizer for future use
```

honing in on promising areas, TPE effectively balances the exploration–exploitation tradeoff—a fundamental challenge in optimization [47]. The ASHA, a resource-efficient variant of Hyperband introduced by Li et al. [48], enhances hyperparameter tuning by allowing early termination of poorly performing trials. This method reallocates resources to more promising configurations without waiting for all trials to complete, significantly improving computational efficiency. ASHA is particularly well-suited for large-scale hyperparameter optimization tasks because it facilitates efficient exploration of the hyperparameter space while ensuring that promising configurations are exploited more effectively over time [48], [49]. This implementation leverages the combined strengths of TPE and ASHA. These well-established methodologies are integrated to manage computational resources more effectively and accelerate model selection. Building on the foundational work of Bergstra et al. [46] and Li et al. [48], we use the TPE's informed sampling method to guide the exploration of promising hyperparameter configurations, while ASHA ensures efficient resource allocation through its early stopping mechanism. The interplay between TPE's probabilistic model and ASHA's bandit-based strategy significantly reduces computational overhead and shortens the time-to-convergence compared to traditional methods. The steps of this approach are outlined in Algorithm 1, where the hyperparameter search space is defined along with budget constraints. The experiments were conducted on NVIDIA DGX A100 system [50]. Key parameters include the following.

1) *CPUs* and *GPUs*: Resources allocated to each trial. In this setup, each trial is allocated 16 CPUs and 0.5 GPUs, ensuring that parallel computations are distributed efficiently.
2) *n_startup_trials*: Set to 64, this parameter defines the number of initial trials that are evaluated before TPE starts using its model to make informed suggestions. This allows for an exploration phase to gather sufficient data for understanding the hyperparameter landscape.
3) *seed*: Set to 42, ensuring reproducibility by initializing random number generators consistently across trials.
4) *n_max_trials*: Limited to 512, serving as a budget constraint, ensuring that the optimization process does not exceed a predefined number of hyperparameter evaluations.
5) *num_epochs*: Defines the maximum number of iterations or epochs that a single trial can run. Set to 4096, ensuring that each model is trained sufficiently without consuming excessive resources.
6) *grace_period*: Set to 16, defining the minimum number of iterations each trial is allowed to run before ASHA evaluates its performance for potential early stopping.
7) *reduction_factor*: Set to 2, this parameter defines the rate at which the number of trials is halved in each successive round of evaluation, concentrating resources on the most promising configurations.

This hybrid optimization strategy has been applied across various benchmark datasets, consistently demonstrating improvements in model accuracy and reductions in computational costs [45]. These results underscore the benefits of combining Bayesian optimization techniques, like TPE, with bandit-based strategies, such as ASHA, to provide a robust solution for hyperparameter optimization. By balancing exploration and exploitation, this approach ensures that the tuning process is efficient and resource-effective.

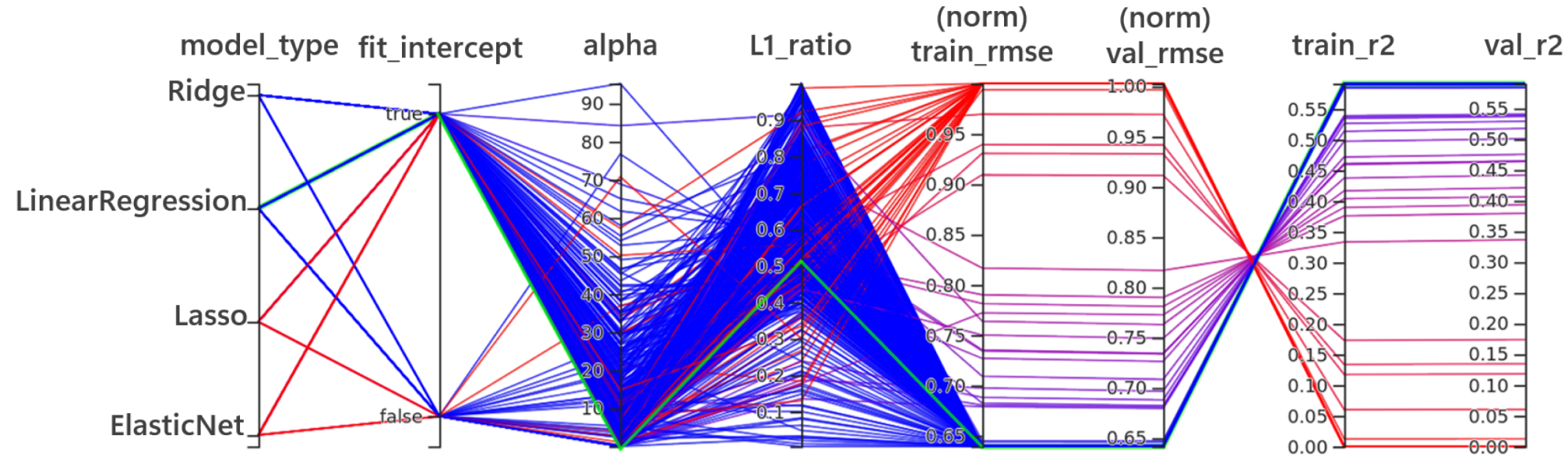

Fig. 5. Parallel coordinates plot for the baseline models, with the top-performing model in terms of validation RMSE is highlighted in green.

TABLE I
Comparison of Model Performance

| Model | | Train | | Validation | | Test† | |
|---|---|---|---|---|---|---|---|
| **Category** | **Name** | **RMSE [dBsm]** | $R^2$ | **RMSE [dBsm]** | $R^2$ | **RMSE [dBsm]** | $R^2$ |
| **Baseline** | **Linear** | 3.2180 | 0.5913 | 3.2298 | 0.5905 | 3.2446 | 0.5904 |
| | **Elastic** | 3.4175 | 0.5390 | 3.4204 | 0.5405 | 3.4375 | 0.5403 |
| | **Lasso** | 3.4353 | 0.5342 | 3.4344 | 0.5370 | 3.4579 | 0.5348 |
| | **Ridge** | 3.2182 | 0.5912 | 3.2302 | 0.5904 | 3.2443 | 0.5905 |
| **Novel** | **DSSP** | 2.5004 | 0.7532 | 2.5748 | 0.7398 | 2.5687 | 0.7433 |

† The Bayesian DSPP captures both epistemic and aleatoric uncertainty, producing 95% per-sample prediction intervals for the test set [Fig. 13(b)]. 2-D histograms of true RCS versus residuals for the baseline and DSPP models (Figs. 7 and 14, respectively) display robust spread metrics–IQR and MAD–providing a fine-grained view of predictive reliability that complements the RMSE and $R^2$ values in Table I. The top-performing baseline and DSPP models, respectively, are highlighted by yellow and green.

## VII. Results and Discussion

### A. Training and Hyperparameter Optimization of the Baseline Models

As discussed in Section IV, the baseline models were chosen for their ability to generalize linear relationships, with each model incorporating different regularization techniques to handle overfitting and multicollinearity, particularly in high-dimensional or correlated data. The implementation of these baseline models was carried out using `scikit-learn` [51] and `PyTorch` [52], with hyperparameter optimization performed via `Optuna` [53], a widely used framework for hyperparameter search and optimization. The search strategy for the optimization follows the method outlined in Algorithm 1, where TPE was used to efficiently explore the hyperparameter space, and ASHA was employed to terminate underperforming trials early, optimizing computational resources. A total of 512 trials were conducted, with performance evaluated based on the training and validation RMSE and $R^2$ scores. The best-performing model for each variant was selected based on the lowest validation RMSE. The training data constituted 70% of the total dataset, while 15% was set aside for validation. The remaining 15% of the data was reserved for testing. The search space for these baseline models was defined as follows.

1) *Model Type (Model):* Linear regression, ridge, lasso, and ElasticNet.
2) *Alpha Parameter ($\alpha$):* Log-uniformly sampled between 0.01 and 100.0 (used for ridge, lasso, and ElasticNet).
3) *Fit Intercept (Intercept):* True, False.
4) *L1 Ratio (Ratio):* Uniformly sampled between 0.0 and 1.0 (only applicable to ElasticNet).

For these models, no epoch-based training was needed, as they directly minimize a regularized objective. A parallel coordinates plot is provided in Fig. 5 to visualize the search space and illustrate how different hyperparameter configurations influenced model performance. The plot depicts that the best-performing baseline model is linear regression with validation $\text{val_RMSE}_{\text{norm}} = 0.6408$ and $\text{val_}R^2 = 0.5905$. The corresponding $\text{val_RMSE}_{\text{unnorm}} = 3.2298$ [dBsm]. The performance metrics for the training, validation, and test data splits, respectively, are summarized in Table I. Among these four baseline variants, the top-performing linear regression (highlighted by yellow in the table) is closely followed by ridge regression. If a grid search was used instead of TPE-ASHA, the search space size would have been much larger due to the need to explore all possible combinations of hyperparameters. The search space would have included the following.

1) *Model Type:* Four choices (linear regression, ridge, lasso, and ElasticNet).
2) *Fit Intercept:* Two choices (True or False).
3) *Alpha:* 100 possible values (log-uniform).
4) *L1 Ratio:* 100 choices (only for ElasticNet).

The total grid search space for each model would be as follows.

1) For *Linear Regression:* Two options (no alpha or L1 ratio involved).
2) For *Ridge and Lasso Regression:* 200 options each (alpha and fit intercept).

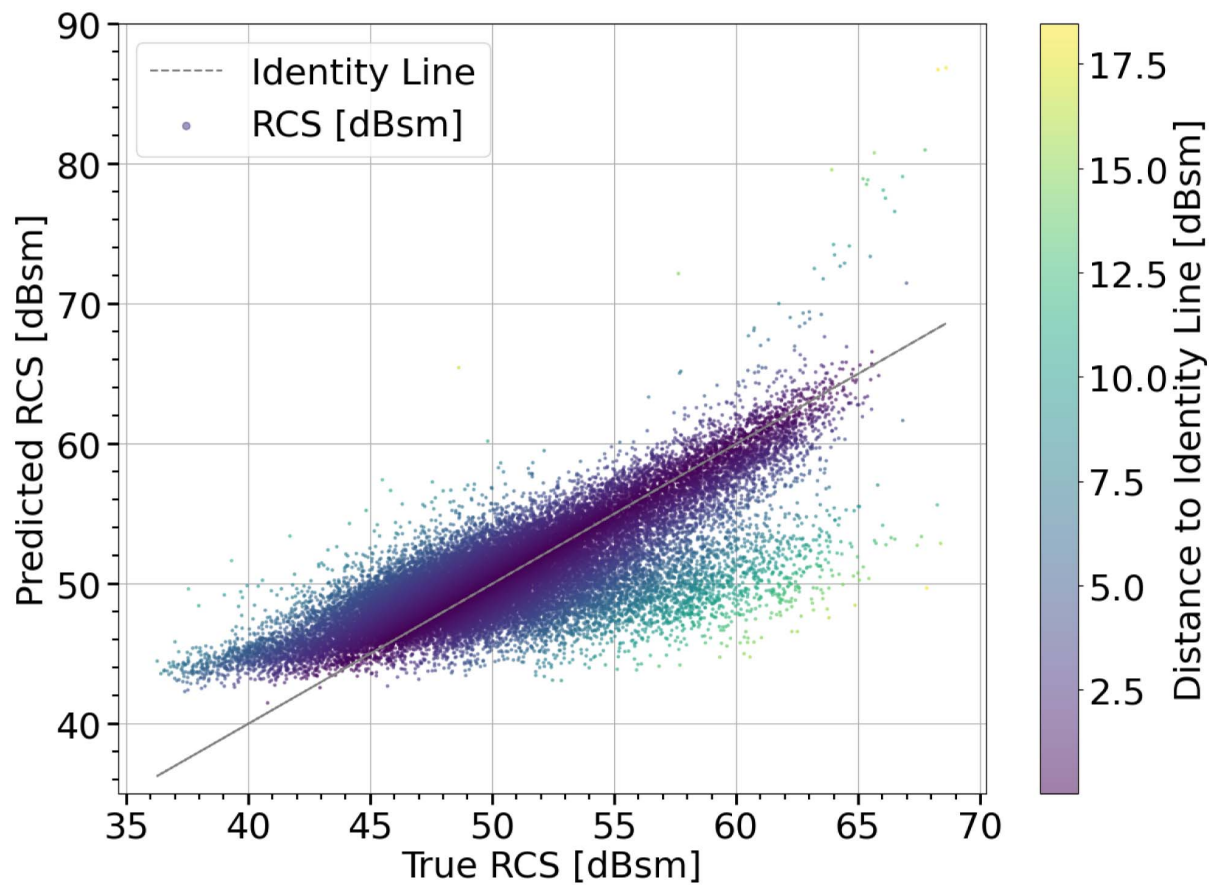

Fig. 6. Scatter plot of predicted versus true RCS for the top-performing baseline (i.e., linear regression) model.

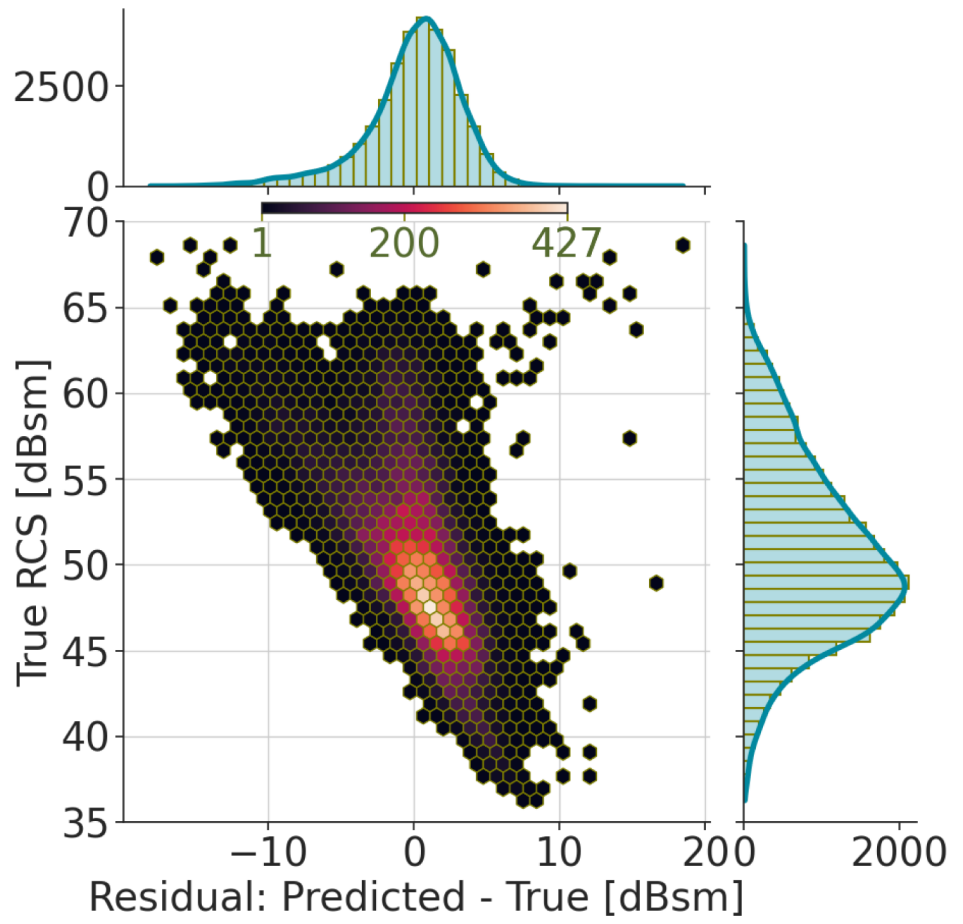

Fig. 7. 2-D histogram plot of true RCS versus residuals for the top-performing baseline (i.e., linear regression) model. Residuals have an IQR=3.6 dBsm and a MAD=1.8 dBsm.

3) For *ElasticNet:* 20 000 options (due to both alpha and L1 ratio).

Fig. 6 provides a scatter plot for predicted versus true RCS [dBsm] pertaining to the test split of the dataset processed by the top-performing linear regression model. Further, Fig. 7 provides a 2-D histogram plot for the true RCS versus residual, along with 1-D histograms for the residual and true RCS. The interquartile range (IQR) and median absolute deviation (MAD) for the residual are 3.6 and 1.8 dBsm, respectively. These baseline models serve as useful benchmarks, illustrating the strengths of regularized linear approaches while allowing comparison with the more complex, uncertainty-aware DSPP model.

## B. Training and Hyperparameter Optimization of the DSPP Model

For training the DSPP model introduced in Section V, we utilized `PyTorch` [52] for model implementation and `GPyTorch` [54] to efficiently handle the GP layers within the DSPP structure. Hyperparameter optimization was carried out using `Optuna` [53], following the method outlined in Algorithm 1. A total of 512 trials were conducted, with performance evaluated based on the training and validation RMSE and $R^2$ scores. The best-performing model was selected based on the lowest validation loss. In DSPP models, validation loss is preferred over validation RMSE when selecting the best model because it captures both the accuracy of the predictions and the model's confidence in those predictions. By using validation loss, we ensure a consistent evaluation with the model's training objective, helping to identify a model that is not only accurate but also well-calibrated in its uncertainty estimates. The training, validation, and testing was based on the same 70%, 15%, and 15% data splits, respectively. The search space for the DSPP model was defined as follows.

1) *Batch Size (B):* 16, 32, 64, 128, 256, 512, 1024, 2048.
2) *Hidden Dimensions (Per Layer) (W):* 2, 3, 4, 5, 6, 7, 10.
3) *Number of Layers (L):* 2, 3, 4, 5, 6, 7.
4) *Number of Inducing Points (M):* 300, 400, 500.
5) *Number of Quadrature Sites (Sigma Points, S):* 6, 8, 10, 12, 15.
6) *Beta Parameter ($\beta$):* 0.01, 0.05, 0.1, 0.2.
7) *Initial Learning Rate ($\alpha_0$):* Uniformly sampled between 0.001 and 0.1.
8) *Optimizer:* Adam, SGD, and RMSprop.
9) *Base Kernel:* RBFKernel, MaternKernel, PeriodicKernel, LinearKernel, PiecewisePolynomialKernel, and RQKernel.
10) *Nu (Smoothness parameter for MaternKernel) ($\nu$):* 0.5, 1.5, 2.5.
11) *Smoothness Parameter (For PiecewisePolynomialKernel) (q):* 0, 1, 2, 3.
12) *Composite Kernel:* True, False.
13) *Composite Kernel Type:* Additive, Multiplicative.
14) *Second Kernel for Composite Type:* Same choices as for the base kernel.

The total search space size for a grid search over the provided hyperparameters can be calculated as follows:

$$\begin{aligned}\text{Total Search Space} = 8 \times 7 \times 6 \times 3 \times 5 \times 4 \times 1000 \\ \times 3 \times 6 \times 3 \times 4 \times 2 \times 2 \times 6. \end{aligned} \tag{26}$$

This would be approximately $1.05 \times 10^{11}$ possible combinations. This vast search space highlights the impracticality of performing grid search for this problem, reinforcing the importance of more efficient search techniques like TPE-ASHA. In addition, the training employed cosine annealing with warm restarts to adjust the learning rate dynamically over time, which has been shown to improve model convergence and performance in neural networks [55]. During the optimization process, several performance metrics were monitored, including RMSE, $R^2$, and loss scores for both training and validation phases. As depicted in Fig. 8, these metrics were visualized using a parallel coordinates plot across all the 512 trials, providing insights into how different hyperparameters impacted the model's performance.

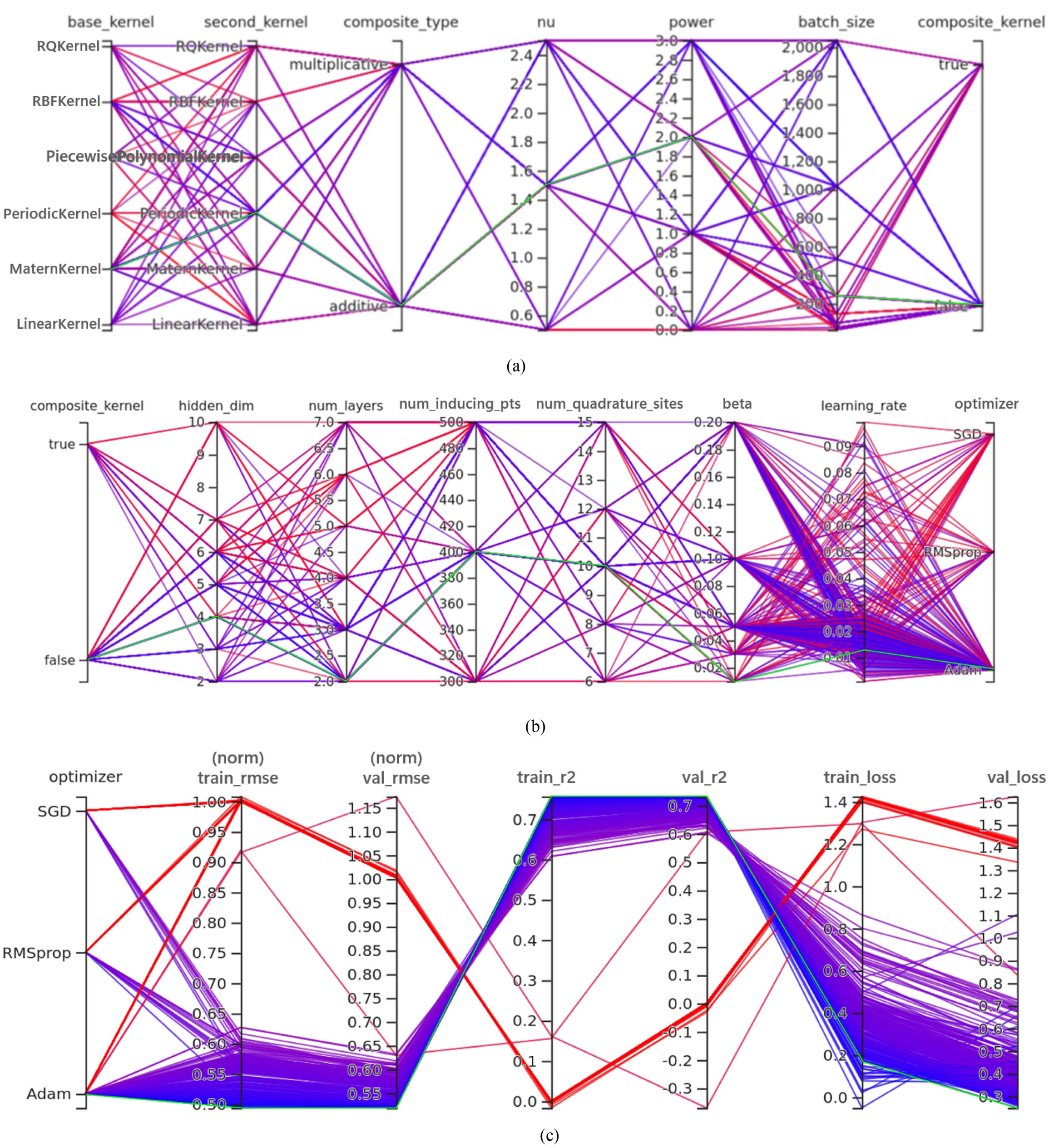

Fig. 8. Parallel coordinates plot for the 512 DSPP trials, with the top-performing model in terms of validation loss is highlighted in green. (a) Plot 1 of 3. (b) Plot 2 of 3. (c) Plot 3 of 3.

The best-performing trial has the lowest validation loss val_loss = 0.2226 at epoch no. 299. Table I lists the RMSE and $R^2$ performance metrics for this best-performing DSPP model in terms of the training, validation, and test data splits, respectively. Figs. 9– 11, respectively, show the loss, RMSE, and $R^2$ plots versus epochs for the best-performing trial. The model architecture for the best-performing DSPP trial is depicted in Fig. 12, with the configuration parameters provided in Table II. Fig. 13 provides a scatter plot for predicted versus true RCS [dBsm] pertaining to the test split of the dataset processed by the top-performing DSPP model, without and with the error bounds estimated by the model, respectively. Further, Fig. 14 provides a 2-D histogram plot for the true RCS versus residual, along with 1-D histograms for the residual and true RCS. For the DSPP's residual, IQR=2 dBsm and MAD=1 dBsm.

## C. Performance Comparison and Observations

Performance of the top DSPP model against the corresponding baseline linear regression model for the same test data split is summarized as follows.

1) *Error Metrics:*

1) The DSPP model demonstrates superior performance compared to the linear regression baseline

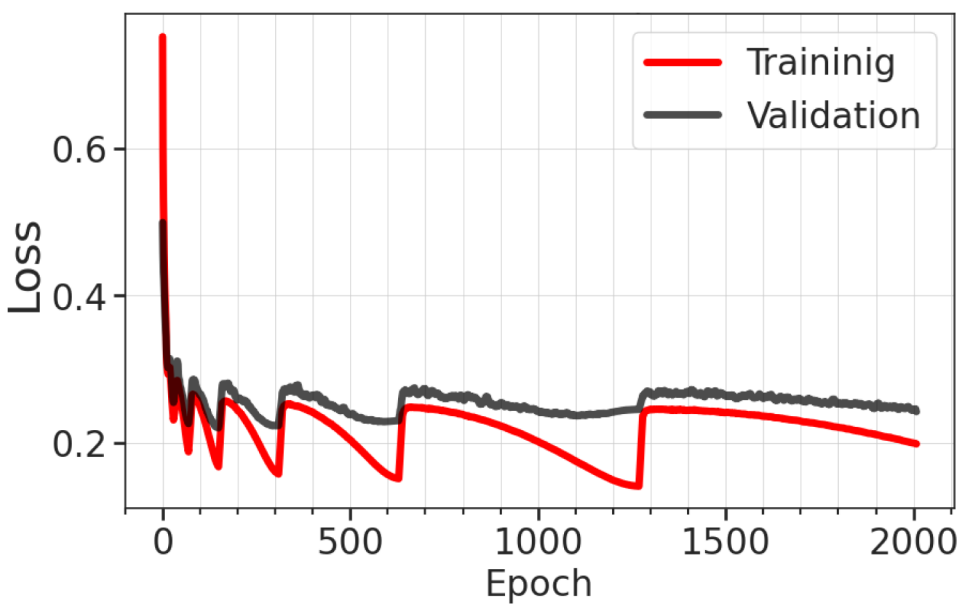

Fig. 9. Loss plots for the top-performing DSPP trial.

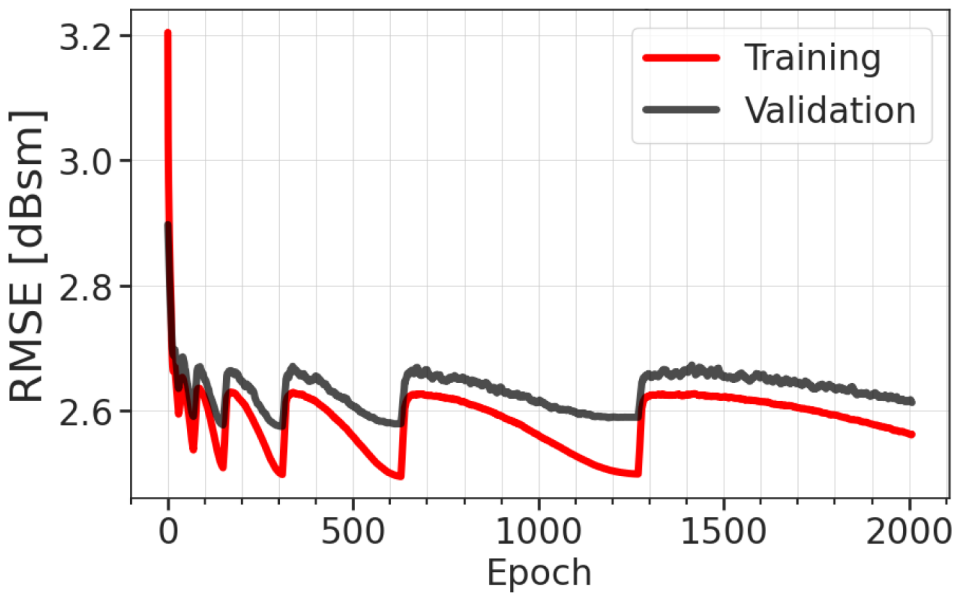

Fig. 10. RMSE plots for the top-performing DSPP trial.

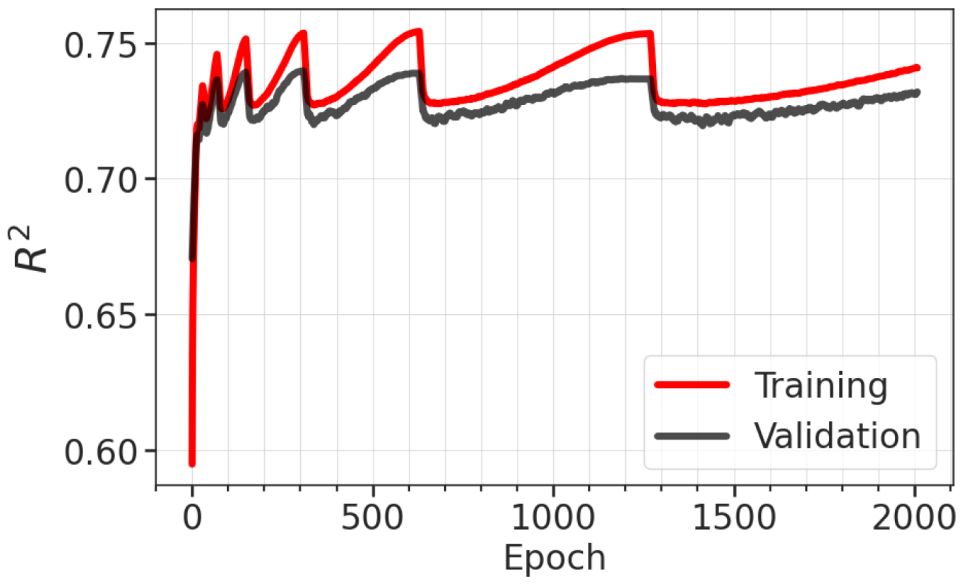

Fig. 11. $R^2$ plots for the top-performing DSPP trial.

(Table I), with a test RMSE of 2.57 dBsm, significantly lower than the baseline linear regression model's RMSE of 3.24 dBsm. This improvement indicates a higher predictive accuracy in the DSPP model.

2) The IQR and MAD values also highlight the DSPP's robustness in error prediction, with an IQR of 2 dBsm and MAD of 1 dBsm (Fig. 7), outperforming the baseline's IQR of 3.6 dBsm and MAD of 1.8 dBsm (Fig. 14). These metrics illustrate that the DSPP model yields more consistent predictions, with fewer large deviations from the true RCS values.

2) *Correlation and Model Fit:*

1) The scatter plots (Figs. 6 and 13) illustrate the relationship between predicted and true RCS values for both models:

1) in the *baseline linear regression plot* (Fig. 6), a broader spread around the identity line suggests more significant errors, particularly at higher RCS values. This observation aligns with the baseline's higher RMSE and lower $R^2$;

TABLE II
Model Configuration for the Top-Performing DSPP Model

| Parameter | Value |
|---|---|
| **General Settings** | |
| Batch Size ($B$) | 256 |
| Beta ($\beta$) | 0.01 |
| Optimizer | Adam |
| Initial Learning Rate ($\alpha_0$) | 0.0118 |
| **Model Architecture** | |
| Hidden Dimension ($W$) | 4 |
| Number of Layers ($L$) | 2 |
| Number of Inducing Points ($M$) | 400 |
| Number of Quadrature Sites ($S$) | 10 |
| **Base Kernel Settings** | |
| Base Kernel | Matérn |
| Nu Parameter ($\nu$) | 1.5 |
| **Composite Kernel Settings** | |
| Composite Kernel | False |

2) the DSPP *plot without error bounds* [Fig. 13(a)] shows a tighter clustering around the identity line, reflecting the DSPP model's better fit and predictive accuracy;

3) the DSPP *plot with error bounds* [Fig. 13(b)] adds valuable interpretability by visualizing the model's confidence in its predictions, with smaller residuals closer to the identity line and error bounds that highlight prediction uncertainty. This feature is an advantage of the DSPP, as it can predict RCS values along with confidence intervals, aiding in its practical utility.

3) *Predictive Power and Explainability:*

1) The DSPP model not only achieves lower RMSE but also demonstrates higher $R^2$ values (0.74 on the test set) compared to the baseline's $R^2$ of 0.59. This improved $R^2$ indicates that the DSPP model explains a larger portion of the variance in RCS, providing stronger predictive insights.

2) The DSPP's error bounds further enhance explainability, offering an estimate of the model's uncertainty, which is essential for applications requiring reliable RCS predictions under varying conditions.

4) *Implications for Model Selection:*

1) The DSPP model's ability to predict error bounds provides an edge over the linear regression baseline, making it a more suitable choice for applications in radar and sensor modeling where both accuracy and reliability are essential.

2) By incorporating both improved predictive performance and uncertainty estimation, the DSPP model addresses limitations observed in simpler models, such as the inability of the linear regression model to handle complex, nonlinear relationships in RCS data effectively.

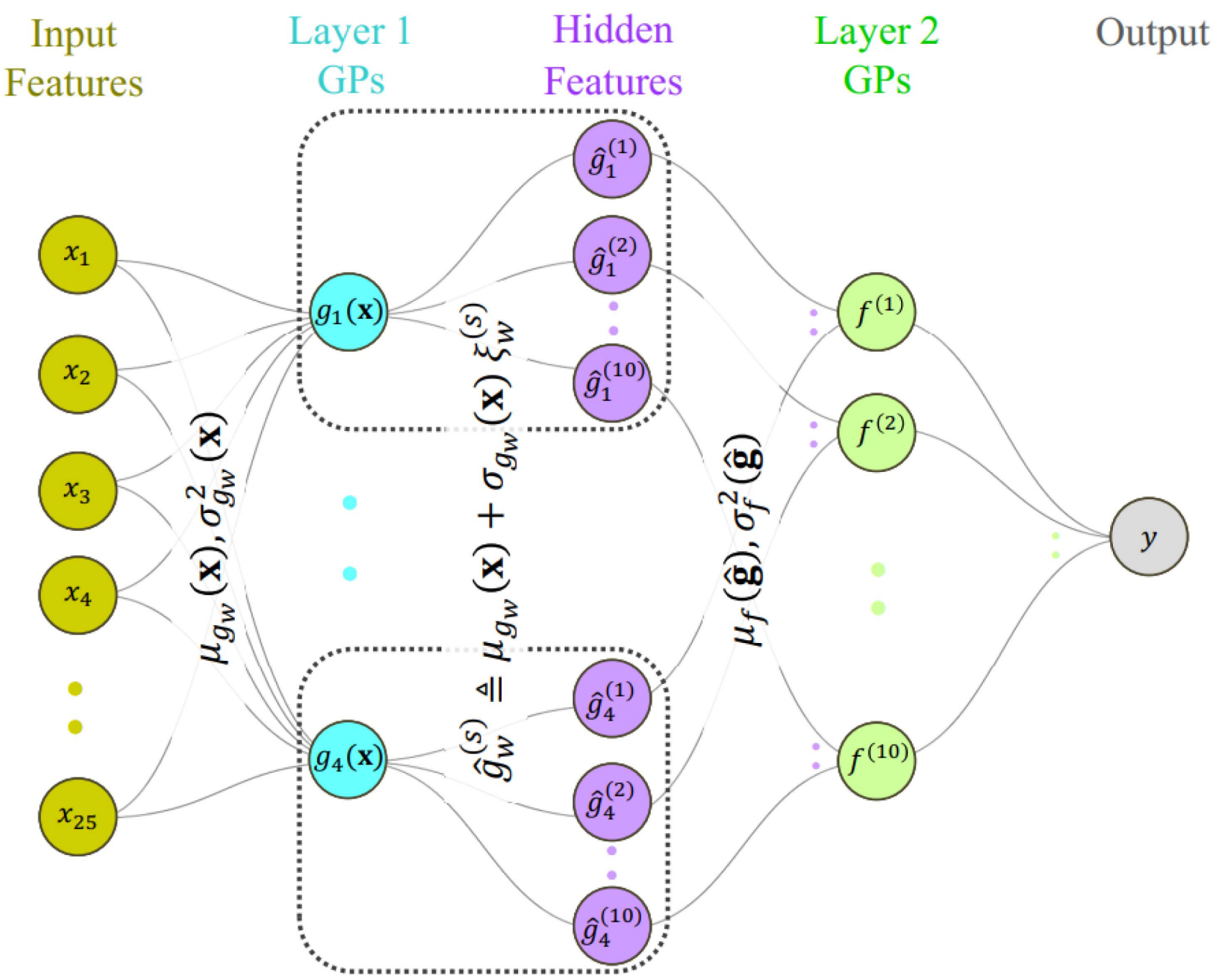

Fig. 12. Architecture of the top-performing novel DSPP model for RCS modeling.

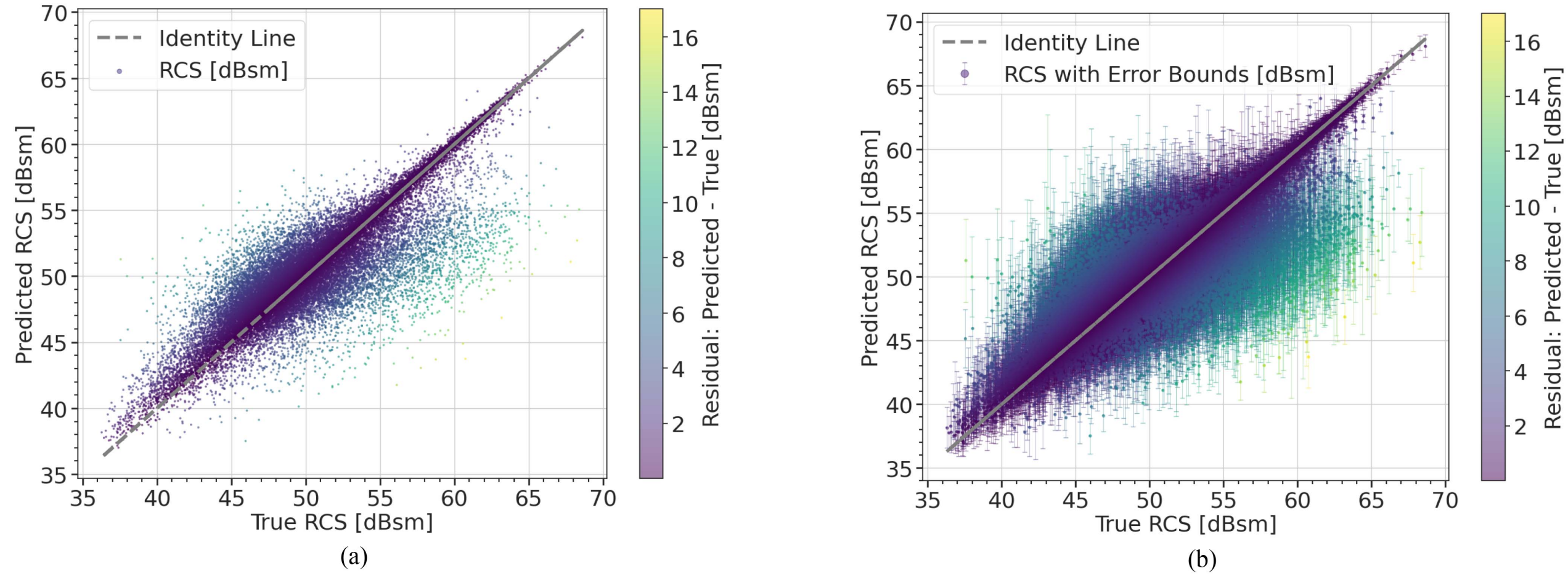

Fig. 13. Scatter plot of predicted versus true RCS for the top-performing DSPP model. (a) Without 95% confidence interval. (b) With 95% confidence interval.

In summary, the DSPP model exhibits clear advantages over the linear regression baseline, with improved accuracy, reliability, and interpretability, as evidenced by lower RMSE and MAD values, higher $R^2$, and the added capability of error bound predictions. These benefits make DSPP a valuable tool for enhanced RCS modeling and reliable decision-making in radar applications.

## D. On Explainability of the DSPP Model

Automatic relevance determination (ARD) is a technique in GP models that assigns individual lengthscales to each input feature, allowing the model to determine their relative importance [56]. For the top-performing DSPP model, ARD operates within the Matérn kernel structure, with dedicated lengthscales for individual features. This setup enables interpretable feature ranking as the model iteratively optimizes these lengthscales alongside other hyperparameters across multiple training epochs, refining them based on each feature's impact on the model's predictive performance.

The Matérn kernel demonstrated superior performance in this study due to its flexibility in capturing complex spatial relationships while controlling the smoothness of the function through the parameter $\nu$. For this work, $\nu = 1.5$ was found to be optimal, striking a balance between modeling smooth yet localized variations in RCS values, particularly those influenced by factors, such as incidence angle and environmental conditions. This aligns with prior studies highlighting the kernel's effectiveness in geospatial and radar-related applications due to its ability to model data with varying degrees of smoothness [2], [3], [57], [58], [59].

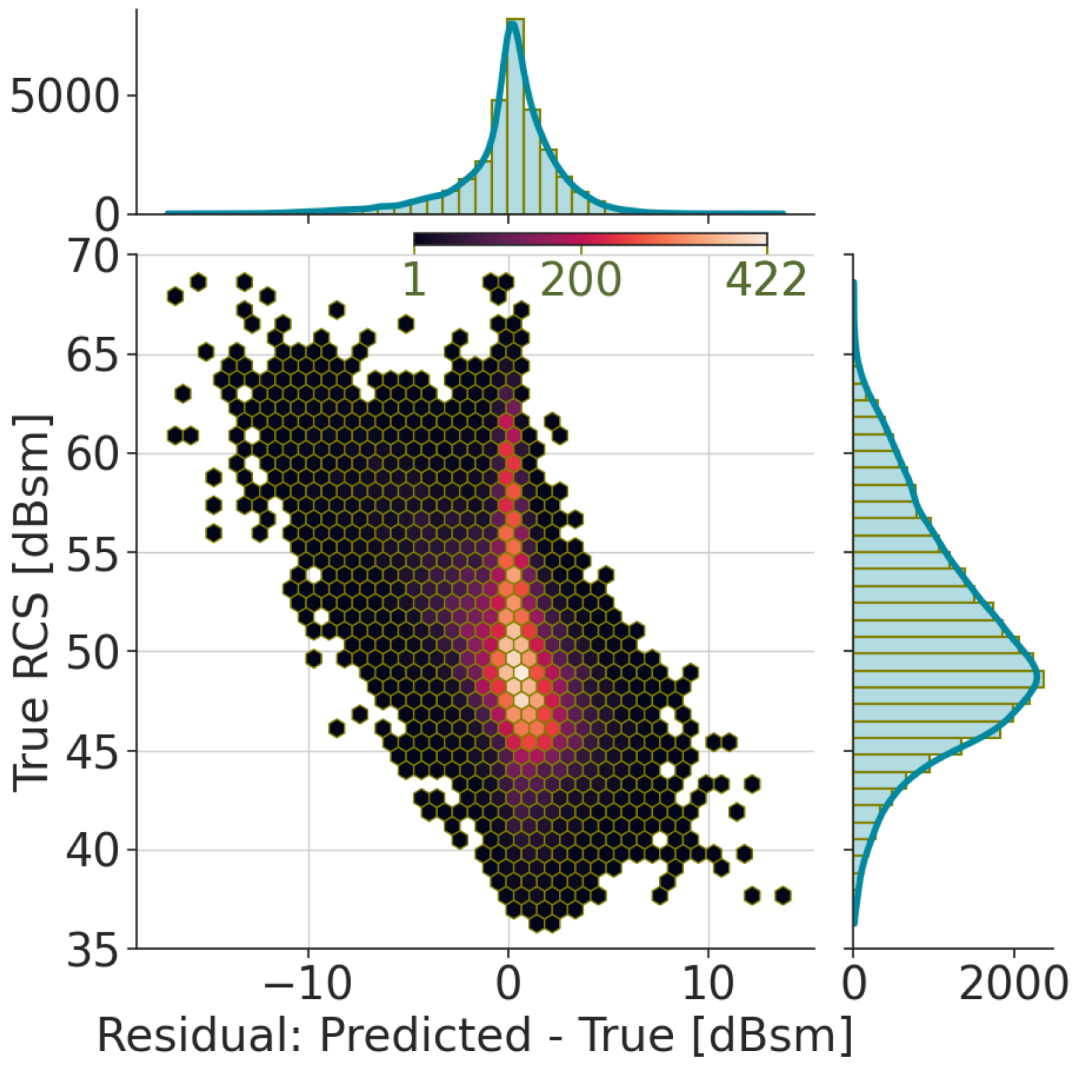

Fig. 14. 2-D histogram plot of true RCS vs. residuals for the top-performing DSPP model. Residuals have an IQR=2 dBsm and a MAD=1 dBsm.

The Matérn kernel with ARD is represented as

$$k_{\text{matern}}(\mathbf{x}, \mathbf{x}') = \sigma^2_{\text{matern}} \frac{2^{1-\nu}}{\Gamma(\nu)} \left( \frac{\sqrt{2\nu}\ \|\mathbf{x} - \mathbf{x}'\|_2}{\ell_i^{\text{matern}}} \right)^{\nu} K_{\nu} \left( \frac{\sqrt{2\nu}\ \|\mathbf{x} - \mathbf{x}'\|_2}{\ell_i^{\text{matern}}} \right) \quad (27)$$

where

1) $\sigma^2_{\text{matern}}$ is a variance parameter controlling the amplitude of the oscillation, learned during training;
2) $\ell_i^{\text{matern}}$ is the lengthscale of the Matérn kernel for a feature $i$, learned during training. Shorter lengthscales imply high sensitivity to nearby points, focusing on localized variations;
3) $\nu$ is the smoothness parameter; for this study, it is set to $\nu = 1.5$;
4) $K_{\nu}$ is the modified Bessel function of the second kind;
5) $\Gamma(\nu)$ is the gamma function.

Upon extensive training, ARD assigns relevance scores to each feature, as observed in the top-performing DSPP model's configuration (see Fig. 15). Feature relevance, "Importance$_i$ %," is calculated by normalizing the inverse of each lengthscale as follows:

$$\text{Importance}_i = \frac{\frac{1}{\ell_i^{\text{matern}}}}{\sum_{j=1}^{25} \frac{1}{\ell_j^{\text{matern}}}} \times 100. \quad (28)$$

Features with shorter lengthscales receive higher relevance scores, indicating greater influence on model predictions. For instance, a feature with a lengthscale ten times smaller than another will possess significantly higher predictive importance due to its increased sensitivity in the DSPP's trained structure.

The Matérn kernel's robustness in balancing predictive accuracy with uncertainty quantification makes it an ideal choice for this DSPP model. Its ability to differentiate features based on their specific contributions to RCS predictions enhances model explainability. ARD provides an interpretable and nuanced view of feature importance that aligns with the DSPP model's architecture and operational behavior.

Each feature in the ARD-based ranking, shown in Fig. 15, is analyzed with reference to its correlation with RCS in Figs. 3–4. Features critical for RCS prediction—such as incidence angle, wind speed, and target dimensions—are accurately captured and ranked by the ARD-enhanced Matérn kernel, providing clear insights into the underlying data structure.

1) *Target Incidence Angle Degrees ($x_{14}$) – 14.12%*:
This feature has the highest ARD score, as expected for RCS modeling, where incidence angle is critical. It directly affects the radar return strength by influencing the scattering properties of the target. The Matérn kernel, particularly with $\nu = 1.5$, captures smooth variations in RCS over small angle changes while accommodating localized effects. The corresponding correlation plot of $\mathbf{x}_{14}$ versus RCS shows a strong dependence of RCS on incidence angle, justifying its high ARD score.
2) *SAR Derived Wind Speed MPS ($x_{22}$) – 9.44%*:
Wind speed impacts sea surface roughness, influencing radar backscatter, particularly for maritime targets. The gradual variation in RCS due to wind speed is effectively modeled by the Matérn kernel, which balances smoothness and flexibility. The $\mathbf{x}_{22}$ versus RCS correlation plot shows significant variation in RCS with wind speed, explaining the high ARD score.
3) *AIS Target Speed MPS ($x_{24}$) – 7.46%*:
The speed of the target influences how it interacts with radar signals. The Matérn kernel, with its ability to assign feature-specific length scales, models these gradual changes efficiently. The $\mathbf{x}_{24}$ versus RCS correlation plot shows a clear dependency of RCS on speed, validating its importance in the ARD ranking.
4) *AIS Target Course Over Ground Degrees ($x_{25}$) – 5.74%*:
The target course over ground benefits from the Matérn kernel's capability to model smooth transitions across varying orientations. Its flexibility, enhanced by ARD, ensures the feature's importance is well-represented. This is confirmed by the corresponding correlation plot for $\mathbf{x}_{25}$ versus RCS, where distinct RCS patterns correspond to different headings.

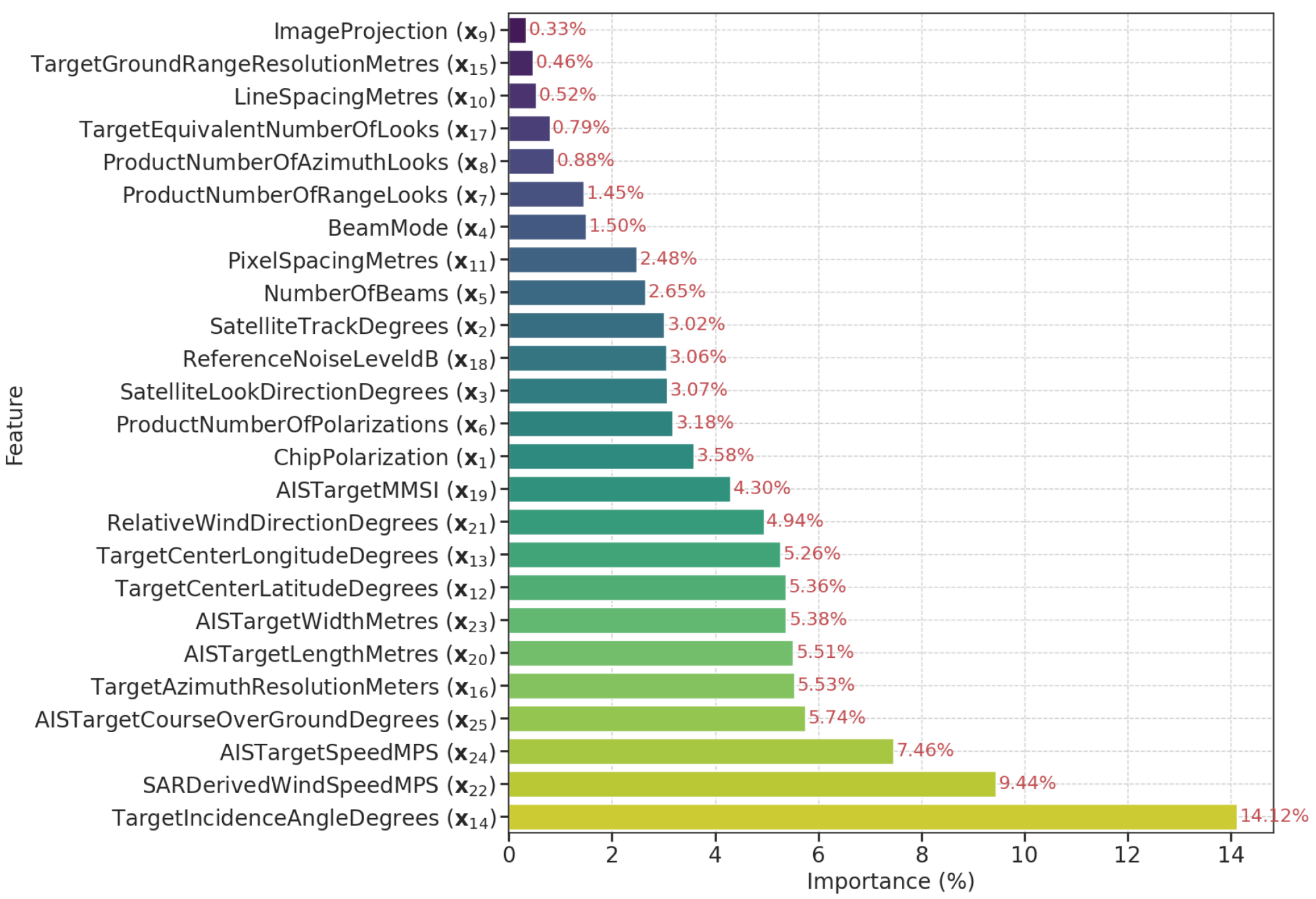

Fig. 15. ARD ranking of the features for the top-performing DSPP model.

5) *Target Azimuth Resolution Meters ($x_{16}$) – 5.53%*:
Azimuth resolution affects the spatial fidelity of radar returns. The Matérn kernel captures gradual variations in resolution effectively, aided by ARD for precise length scale adjustments. The $\mathbf{x}_{16}$ versus RCS correlation plot illustrates how higher azimuth resolution improves RCS detail, reflected in the model's high ranking for this feature.
6) *AIS Target Length Metres ($x_{20}$) – 5.51%*:
Though traditionally one of the most critical features for RCS, it ranks slightly lower here, indicating the added predictive value of other parameters. The Matérn kernel, with ARD, captures the smooth variations across target sizes while tailoring the length scale to this feature. The corresponding correlation plot for $\mathbf{x}_{20}$ versus RCS shows a clear relationship between target length and RCS, supporting the ARD score.
7) *AIS Target Width Metres ($x_{23}$) – 5.38%*:
Width impacts the radar cross-sectional area, contributing to RCS prediction. The Matérn kernel smooths the influence of width while ARD ensures feature-specific modeling. The $\mathbf{x}_{23}$ versus RCS correlation plot affirms its relevance in defining the physical target dimensions.
8) *Target Center Latitude Degrees ($x_{12}$) – 5.36%*:
Latitude offers spatial context that indirectly impacts RCS through environmental factors. The Matérn kernel models gradual geographical effects effectively, while ARD allows fine-grained adjustments for feature relevance. This is confirmed by the $\mathbf{x}_{12}$ versus RCS correlation plot.
9) *Target Center Longitude Degrees ($x_{13}$) – 5.26%*:
Similar to latitude, longitude provides essential spatial context. Together with latitude, it influences environmental conditions affecting radar backscatter. The Matérn kernel and ARD capture these nuanced interactions, evident in the $\mathbf{x}_{13}$ versus RCS correlation plot.
10) *Relative Wind Direction Degrees ($x_{21}$) – 4.94%*:
Relative wind direction, combined with wind speed, affects sea surface roughness and radar backscatter. The Matérn kernel, augmented with ARD, captures gradual environmental impacts effectively. The $\mathbf{x}_{21}$ versus RCS correlation plot highlights this feature's importance in maritime RCS prediction.
11) *AIS Target MMSI ($x_{19}$) – 4.30%*:
The MMSI identifier serves a dual purpose in this study. First, it correlates with specific vessel types or behaviors, indirectly affecting RCS. The Matérn kernel effectively captures these implicit relationships, with ARD adjusting relevance to reflect their impact. Second, the MMSI provides a mechanism for tracking repeated acquisitions of the same vessel, enabling insights into temporal variations in radar returns. This dual functionality contributes to its moderate importance in the ARD ranking, as evidenced by the feature's influence on the model's predictions and the corresponding correlation plot for $\mathbf{x}_{19}$ versus RCS.
12) *Chip Polarization ($x_{1}$) – 3.58%*:
Polarization affects radar signal interaction with the target. The Matérn kernel models smooth transitions while ARD ensures feature-specific tuning. The correlation plot for $\mathbf{x}_1$ versus RCS demonstrates distinct RCS variations based on polarization.

13) *Product Number of Polarizations ($x_6$) – 3.18%*:
This feature indicates the diversity of polarizations in radar measurements, providing additional information on target structure. The Matérn kernel smooths the variations in RCS, with ARD ensuring accurate relevance ranking, as shown in the $\mathbf{x}_6$ versus RCS correlation plot.
14) *Satellite Look Direction Degrees ($x_3$) – 3.07%*:
Look direction affects the aspect angle, with the Matérn kernel modeling smooth directional transitions. ARD refines its length scale for feature-specific importance. The $\mathbf{x}_3$ versus RCS correlation plot supports the relevance of this feature.
15) *Reference Noise Level dB ($x_{18}$) – 3.06%*:
This feature calibrates signal-to-noise in RCS predictions, especially in low-signal conditions. The Matérn kernel captures this effect effectively, as reflected in the $\mathbf{x}_{18}$ versus RCS correlation plot.
16) *Satellite Track Degrees ($x_2$) – 3.02%*:
Track angle subtly impacts radar-target orientation, captured by the Matérn kernel. ARD ensures this feature's relevance is properly modeled. The $\mathbf{x}_2$ versus RCS correlation plot suggests a minor influence on RCS, consistent with the ARD score.
17) *Number of Beams ($x_5$) – 2.65%*:
The number of beams affects radar coverage. The $\mathbf{x}_5$ versus RCS correlation plot shows limited impact, explaining the lower ARD score.
18) *Pixel Spacing Metres ($x_{11}$) – 2.48%*:
Pixel spacing impacts image resolution, indirectly affecting RCS. The Matérn kernel smooths this feature's impact, as confirmed by the $\mathbf{x}_{11}$ versus RCS correlation plot.
19) *Beam Mode ($x_4$) – 1.50%*:
Beam mode affects radar coverage pattern but has limited influence on RCS. This low relevance is consistent with the $\mathbf{x}_4$ versus RCS correlation plot.
20) *Product Number of Range Looks ($x_7$) – 1.45%*:
Range looks add data redundancy, with minimal impact on RCS, as shown in the $\mathbf{x}_7$ versus RCS correlation plot.
21) *Product Number of Azimuth Looks ($x_8$) – 0.88%*:
Similar to range looks, azimuth looks hold limited relevance, with low impact confirmed in the $\mathbf{x}_8$ versus RCS correlation plot.
22) *Target Equivalent Number of Looks ($x_{17}$) – 0.79%*:
This measure of data robustness has limited impact on RCS prediction, as reflected in the $\mathbf{x}_{17}$ versus RCS correlation plot.
23) *Line Spacing Metres ($x_{10}$) – 0.52%*:
Line spacing affects SAR image quality but has minimal relevance for RCS, as confirmed in the $\mathbf{x}_{10}$ versus RCS correlation plot.
24) *Target Ground Range Resolution Metres ($x_{15}$) – 0.46%*:
Ground range resolution provides spatial detail but is secondary in RCS modeling. This is reflected in the $\mathbf{x}_{15}$ versus RCS correlation plot and ARD score.
25) *Image Projection ($x_9$) – 0.33%*:
This feature has a constant value (ground-range) across the dataset, rendering it noninformative for RCS prediction. The $\mathbf{x}_9$ versus RCS correlation plot confirms this minimum impact.

In summary, the Matérn kernel with ARD demonstrates remarkable flexibility and capability in modeling the diverse relationships between input features and RCS. Its ability to assign unique length scales per feature ensures both interpretability and precision in feature ranking. Validation from the ARD-based ranking (Fig. 15) and RCS correlation plots (Figs. 3 and 4) underscores its effectiveness in this application. Because the 25 input features are treated as design variables, in testing scenarios where some of these features are unavailable, the missing parameters can be held at their nominal values or the DSPP model can be fine-tuned or retrained on the remaining inputs. The ARD importance ranking in Fig. 15 identifies the most critical features, ensuring these simplifications incur minimal loss of accuracy.

## VIII. CONCLUSION

In this study, we have developed a novel DSPP model for RCS prediction in spaceborne SAR imagery, using a dataset of AIS-verified RADARSAT-2 ship detections. By leveraging the DSPP model's Bayesian framework, we achieved not only improved prediction accuracy but also quantifiable error bounds, addressing key limitations of traditional regression approaches in handling uncertainty and complex, nonlinear relationships within RCS data.

Our results demonstrate that the DSPP model consistently outperformed traditional multiple linear regression variants, as evidenced by lower RMSE values and higher $R^2$ scores across training, validation, and test splits. The DSPP's use of the Matérn kernel was instrumental in capturing intricate interactions among the 25 features. Each node in the DSPP architecture employs its own kernel, with parameters, such as feature-specific lengthscales independently learned through ARD. This nodewise tuning allows the model to adaptively capture hierarchical relationships, with lower layers focusing on broader patterns and higher layers refining representations to account for localized, complex feature interactions. Features with shorter learned lengthscales emerged as more influential, highlighting their sensitivity to fine-grained variations in the data. This architecture enabled nuanced modeling of both gradual environmental changes, such as radar-derived wind speed, and abrupt variations, such as ship-specific parameters, resulting in a more robust and interpretable predictive framework.

The model's ARD feature ranking underscored the importance of incidence angle, wind speed, and target-specific

AIS parameters, reaffirming these as critical variables in RCS prediction. This interpretability, facilitated by ARD within the DSPP framework, aligns well with XAI principles, making the model particularly suited for operational deployment in defence and maritime applications where transparency in predictions is vital.

Looking forward, several avenues exist for extending this work. First, incorporating additional datasets from sensors with similar C-band center frequencies, such as RADARSAT Constellation Mission and Sentinel-1, would enhance the model's cross-sensor robustness and applicability across broader operational scenarios. To enable this cross-sensor generalization, the sensor's name will be included as an explicit feature, allowing the model to account for potential variations in sensor characteristics and data properties. In addition, including data from sensors with other center frequencies, such as X-band or L-band, would further extend the model's generalization capacity, with the center frequency used as an explicit feature to capture frequency-dependent RCS variations. Second, addressing data imbalances, particularly in ship type diversity, will be crucial in refining the model's generalizability. Employing balanced sampling or data augmentation strategies could mitigate the model's sensitivity to overrepresented classes and improve performance across diverse maritime targets.

Finally, an exciting direction for future research is the potential to reverse the RCS-prediction framework to estimate key maritime parameters, such as the ship's radial speed, course over ground, and environmental variables (e.g., wind speed and direction). By leveraging the model's sensitivity to these features, we aim to explore how the DSPP model could support inverse predictions for improved situational awareness and decision-making in maritime surveillance.

Through these contributions, our DSPP-based approach advances the state of RCS modeling by offering a powerful, interpretable, and flexible model for accurate RCS predictions in complex radar data, laying the groundwork for enhanced SAR-based ship detection, tracking, and classification applications.

## ACKNOWLEDGMENT

The authors gratefully acknowledge Dr. Jessica M. Topple of the DRDC – Atlantic Research Centre for her careful reading of the manuscript and for providing helpful comments that improved its clarity.

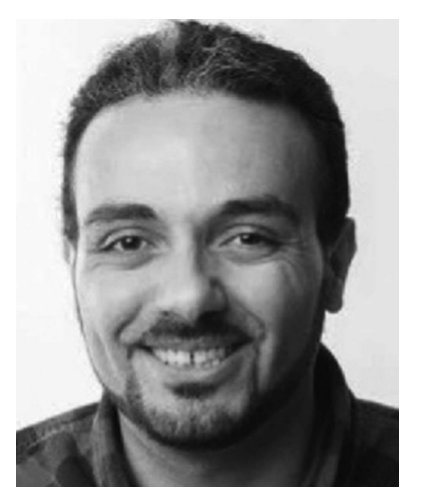

**Khalid El-Darymli** (Senior Member, received the B.Sc. degree in electrical engineering from Garyounis University, Benghazi, Libya, in 2001, the M.Sc. degree in computer and information engineering from the International Islamic University of Malaysia, Kuala Lumpur, Malaysia, in 2006, and the Ph.D. degree (with distinction) in electrical engineering from Memorial University, St. John's, NL, Canada, in 2015.

From 2010 to 2014, he was a Doctoral Researcher with C-CORE, St. John's, NL, Canada, where he developed algorithms for target recognition in SAR imagery. From 2014 to 2017, he worked as a Senior Engineer with Northern Radar Inc., St. John's, NL, Canada, codeveloping a high-frequency software-defined radar for coastal ocean applications. Concurrently, he served as an instructor for both graduate and undergraduate courses in antennas with the Department of Electrical Engineering, Faculty of Engineering and Applied Science, Memorial University, St. John's, NL, Canada, where he was appointed Adjunct Professor with the Department of Electrical Engineering, Faculty of Engineering and Applied Science, in 2025, and is currently a Fellow of the School of Graduate Studies. From mid-2017 to early 2021, he was a Research and Development Scientist with MDA Systems, Richmond, BC, Canada, where he contributed to several projects, including laser guide stars for optical communications, ground moving target indication, velocity estimation in SAR imagery, and target recognition in inverse SAR imagery. Since mid-2021, he has been a Defence Scientist with Defence Research and Development Canada (DRDC), Ottawa, ON, Canada, focusing on various aspects of SAR and ISAR imagery, along with other research and development activities. In June 2023, he was appointed Affiliate Faculty with the Fowler School of Engineering, Chapman University, Orange, CA, USA.

Dr. El-Darymli is a licensed Professional Engineer (P.Eng.) with the Association of Professional Engineers and Geoscientists of Alberta. He was the recipient of the Ocean Industries Student Research Award from the Research and Development Corporation (InnovateNL), Government of Newfoundland and Labrador, for his work on SAR Detection in Cluttered Environments during his Ph.D. studies. He was also the recipient of the Annual Newfoundland Electrical and Computer Engineering Conference 2013 Wally Read Best Student Paper Award. During his M.Sc. studies, he developed an award-winning Speech to American Sign Language Interpreter System using machine learning techniques.

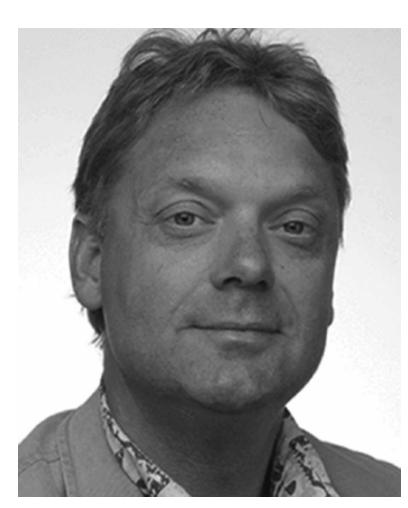

**Christoph H. Gierull** (Senior Member, received the Dr.-Ing. degree in electrical engineering from Ruhr-Universität Bochum, Bochum, Germany, in 1995.

From 1991 to 1994, he was a Scientist with the Research Establishment for Applied Science, Wachtberg, Germany. In 1994, he joined the German Aerospace Center, Oberpfaffenhofen, Germany, where he led the SAR Simulation Group. As part of a DLR team, he assured the X-band interferometric SAR performance during the Space-Shuttle Radar Topography Mission with NASA Mission Control, Houston, TX, USA. From 2006 to 2009, he was with Fraunhofer FHR, Wachtberg, Germany. Since 2000, he has been a Senior Defence Scientist with Defence Research and Development Canada, Ottawa, ON, Canada. In 2010, he became the Leader of the Space-Based Radar Group. In 2011, he was appointed as an Adjunct Professor with Université Laval, Quebec City, QC, Canada, and since 2017, he has been also with Simon Fraser University, Burnaby, BC, Canada. He has authored or coauthored numerous journal publications, scientific reports, and a chapter in *Applications of Space-Time Adaptive Processing* (IEE, 2004). He has coedited and authored a chapter in, *Novel Radar Techniques and Applications* (IET, 2017). He filed several Report-of-Inventions and was granted a CAN/U.S. patent on vessel detection in SAR imagery.

Dr. Gierull was elected as a Fellow of IET. He was the recipient of the Best Annual Paper Award of the Association of German Electrical Engineers in 1998, and the co-recipient of the best paper awards at the International Radar Conference 2004, and the EUSAR 2006, as well as EUSAR 2016. He was also the recipient of DRDC's highest S&T Performance Excellence Award three times, in 2013 and 2020, and again in 2022 in addition to the Geoscience and Remote Sensing Society 2016 J-STARS Paper Award. He and his team were the recipients of the Square Dance Arnold Award in 2021. He initiated and co-organized the first Special Issue on Multichannel Space-Based SAR in the Journal of Selected Topics in Applied Earth Observations and Remote Sensing in 2015. He was an Associate Editor (AE) of EURASIP's Signal Processing in 2004. Since 2019, has been serving as an AE for Transactions on Geoscience and Remote Sensing. He has been a Technical Advisor to the Canadian Space Agency on several SAR missions including RADARSAT-2, RCM, and its follow-on. In 2021, the European Space Agency selected him as a Member on its Sentinel-1 Next-Generation Mission Advisory Group.

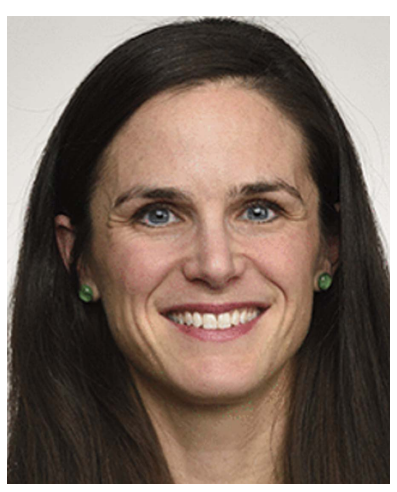

**Katerina Biron** received the B.Sc. degree in mechanical engineering from Old Dominion University in Virginia, Norfolk, VA, USA, 2007, and the M.Sc. and Ph.D. degrees in electrical engineering and signal processing from the University of New Brunswick, Fredericton, NB, Canada, in 2010 and 2017.

She currently is a Research Scientist with the Space and Intelligence, Surveillance and Reconnaissance (ISR) Applications Section, Defence Research and Development Canada, Ottawa Research Centre, Ottawa, ON, Canada. Her research interests include machine learning and Big Data analysis, data processing and exploitation, and geospatial data management, for remote sensing and space-based ISR for defence and civilian applications.

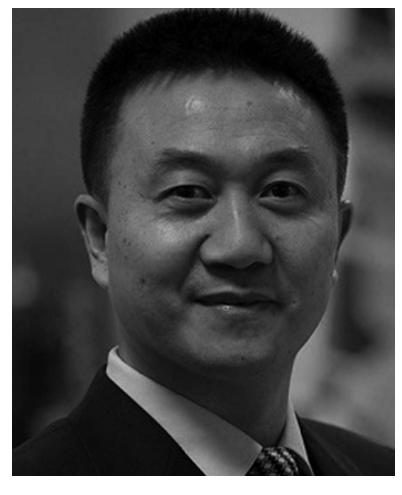

**Weimin Huang** (Senior Member, received the B.S., M.S., and Ph.D. degrees in radio physics from Wuhan University, Wuhan, China, in 1995, 1997, and 2001, respectively, and the M.Eng. degree in electrical engineering from the Memorial University of Newfoundland, St. John's, NL, Canada, in 2004.

From 2008 to 2010, he was a Design Engineer with Rutter Technologies, St. John's, NL, Canada. Since 2010, he has been with the Faculty of Engineering and Applied Science, Memorial University of Newfoundland, where he is currently a Professor. He has authored or coauthored more than 300 refereed research articles. He has edited the book *Ocean Remote Sensing Technologies: High Frequency, Marine, and GNSS-Based Radar*. His research interests include mapping of oceanic surface parameters via high-frequency ground wave radar, X-band marine radar, synthetic aperture radar, and global navigation satellite systems.

Dr. Huang was a Member and the Co-Chair of the Electrical and Computer Engineering Evaluation Group for Natural Sciences and Engineering Research Council of Canada Discovery Grants, from 2018 to 2021. He was the recipient of a Postdoctoral Fellowship from the Memorial University of Newfoundland, the Discovery Accelerator Supplements Award from NSERC in 2017, and Geoscience and Remote Sensing Society 2019 Letters Prize Paper Award as well as some other teaching and research awards. He is an Oceanic Engineering Society (OES) Distinguished Lecturer. He is also an Area Editor for Canadian Journal of Electrical and Computer Engineering, an Associate Editor for Transactions on Geoscience and Remote Sensing, Geoscience and Remote Sensing Letters, Journal of Oceanic Engineering, *Remote Sensing*, *Frontiers in Marine Science*, *Frontiers in Remote Sensing*, and has been a Guest Editor for Journal of Selected Topics in Applied Earth Observations and Remote Sensing and six other journals. He has been a Reviewer for more than 120 international journals and a reviewer for many international conferences. He has been a Technical Program Committee Member. He is the General Co-Chair and the Technical Program Co-Chair for the 20th International Symposium on Antenna Technology and Applied Electromagnetics, and OES 13th Currents, Waves, and Turbulence Measurement Workshop.